\documentclass[useAMS,usenatbib]{mnras}

\usepackage{graphicx}
\usepackage{amsmath}
\usepackage{natbib}
\usepackage{color}
\usepackage[T1]{fontenc}

\newcommand*\msolar{\mathop{}\!\,\mathcal{M}_{\sun}}

\newcommand*\kmps{\mathop{}\!\,\mathrm{km\,s}^{-1}}
\newcommand*\gyr{\mathop{}\!\,\mathrm{Gyr}}
\newcommand*\magnitude{\mathop{}\!\,\mathrm{mag}}

\title[The WDLFs from the PS1 3$\upi$ Survey]{The White Dwarf Luminosity Functions from the Pan--STARRS 1 3$\upi$ Steradian Survey}
\author[M. C. Lam et al.]{Marco~C.~Lam\thanks{E-mail: c.y.lam@ljmu.ac.uk}$^{1,2}$,
Nigel~C.~Hambly$^{2}$,
Nicholas~Rowell$^{2}$,
Kenneth~C.~Chambers$^{3}$,
\newauthor
Bertrand~Goldman$^{4}$,
Klaus~W.~Hodapp$^{3}$,
Nick~Kaiser$^{3}$,
Rolf-Peter~Kudritzki$^{3}$,
\newauthor
Eugene~A.~Magnier$^{3}$,
John~L.~Tonry$^{3}$,
Richard~J.~Wainscoat$^{3}$ and
\newauthor
Christopher~Waters$^{3}$
\\
\\
$^{1}$Astrophysics Research Institute, Liverpool John Moores University, IC2, LSP, 146 Brownlow Hill, Liverpool L3~5RF, UK\\ 
$^{2}$Institute for Astronomy, University of Edinburgh, Royal Observatory of Edinburgh, Blackford Hill, Edinburgh EH9~3HJ, UK\\
$^{3}$Institute for Astronomy, University of Hawaii at Manoa, Honolulu, HI 96822, USA\\
$^{4}$Max-Planck-Institut f\"ur Astronomie, K\"onigstuhl 17, D-69117 Heidelberg, Germany
}

\begin{document}

\date{original form 6 September 2018}

\pagerange{\pageref{firstpage}--\pageref{lastpage}} \pubyear{2018}

\maketitle

\label{firstpage}

\begin{abstract}

A large sample of white dwarfs is selected by both proper motion and colours from the Pan-STARRS 1 $3\upi$ Steradian Survey Processing Version 2 to construct the White Dwarf Luminosity Functions of the discs and halo in the solar neighbourhood. Four-parameter astrometric solutions were recomputed from the epoch data. The generalised maximum volume method is then used to calculate the density of the populations. After removal of crowded areas near the Galactic plane and centre, the final sky area used by this work is $7.833$\,sr, which is $83\%$ of the $3\upi$ sky and $62\%$ of the whole sky. By dividing the sky using Voronoi tessellation, photometric and astrometric uncertainties are recomputed at each step of the integration to improve the accuracy of the maximum volume. Interstellar reddening is considered throughout the work. We find a disc-to-halo white dwarf ratio of about $100$.

\end{abstract}

\begin{keywords}
proper motions -- surveys -- stars: luminosity function, mass function -- white dwarfs -- solar neighbourhood.
\end{keywords}

\section{Introduction}

Main sequence~(MS) stars with initial mass less than $8\msolar$ end up as white
dwarfs~(WDs) at the end of their lives. Since this mass range 
encompasses the vast majority of stars in the Galaxy, these degenerate remnants
are the most common final product of stellar evolution. In this state there is
little nuclear burning to replenish the energy they radiate away. As a consequence, the
luminosity and temperature decrease monotonically with time. The electron
degenerate nature means that a WD with a typical mass of $0.6\msolar$ has a
similar size to the Earth which gives rise to their high densities, large
surface gravities and low luminosities. The coolest WDs in particular have
neutral colours and very low luminosities and are consequently very hard to
study.

The use of the white dwarf luminosity function~(WDLF) as cosmochronometer was
first introduced by \citet{1959ApJ...129..243S}. Given a finite age of the
Galaxy, there is a minimum temperature below which no white dwarfs can reach in
a limited cooling time. This limit translates to an abrupt downturn in the WDLF
at faint magnitudes. Evidence of such behaviour was observed by
\citet{1979ApJ...233..226L}, however, it was not clear at the time whether it
was due to incompleteness in the observations or to some defect in the
theory~(e.g.,~\citealp{1984ApJ...282..615I}). A decade later,
\citet{1987ApJ...315L..77W} gathered concrete evidence for the downturn and
estimated the age\footnote{The ``age'' refers to the total time since the oldest
WD progenitor arrived at the zero-age main sequence.} of the disc to be $9.3 \pm
2.0\gyr$~(see also \citealt{1988ApJ...332..891L}). While most studies focused on
the Galactic discs~\citep{1989LNP...328...15L, 1992ApJ...386..539W, 1995LNP...443...24O,
1998ApJ...497..294L, 1999MNRAS.306..736K, 2012ApJS..199...29G}, some worked with
open clusters~\citep{2000ApJ...529..318R}, globular
clusters~\citep{2002ApJ...574L.155H,2009ApJ...705..408K,2010ApJ...708L..32B},
the stellar halo~(\citealt{2006AJ....131..571H}, hereafter H06; \citealt{2011MNRAS.417...93R}, hereafter RH11; \citealt{2017AJ....153...10M}, hereafter M17) and the Galactic bulge~\citep{2015ApJ...810....8C}. 

Algorithms for recovering the age and star formation history~(SFH) of a stellar population from the WDLF have also been developed~(\citealp{1990ApJ...352..605N}; 
\citealp{2001ASPC..245..328I}; \citealp{2013MNRAS.434.1549R}, hereafter R13).
For example, a short burst of increased star formation would appear as a bump in
the WDLF. The use of WDLF inversion to derive the SFH is still in its infancy.
R13 developed an inversion algorithm that requires input WDLF and WD atmosphere
evolution models, and is similar to other inversion algorithms applied on
colour-magnitude diagrams. However, there is some debate over the smoothing and
possible amplification of noise during the application of Richardson-Lucy
deconvolution~\citep{1972JOSA...62...55R, 1974AJ.....79..745L} and the
determination of the point of convergence. \citet{2014ApJ...791...92T} used a
set of confirmed spectroscopic WDs with well determined distance, temperature
and surface gravity, hence the mass and radius, to derive the age of each
individual WD. In their case, the derived SFH was mostly consistent with R13 but
it lacks a peak at recent times which they claim as noise being
amplified by the algorithm developed by R13. Overall, the results are broadly
consistent with each other as well as those derived from the inversion of
colour-magnitude diagrams with different algorithms~(\citealp{2002A&A...390..917V};
\citealp{2006A&A...459..783C}).

Hot WDs have UV excess compared to the MS stars. However, warm and cool WDs
overlap with the MS stars in any colour combination so it is difficult to
distinguish them in colour-colour space. In the ultracool regime, 
collisonally--induced absorption due to molecular hydrogen (H2CIA)
makes them blue and so they deviate from MS colours; however, they are
intrinsically too faint to be found in most surveys. To date, there are 19,712
WDs in the catalogue of spectroscopically confirmed isolated WD from the Sloan
Digital Sky Survey~(SDSS) DR7~\citep{2013ApJS..204....5K} and an addition of
8,441 and 3,671 from SDSS DR10 and DR12 respectively~\citep{2015MNRAS.446.4078K,
2016MNRAS.455.3413K}. A lot of them are either false positives of follow up
observations targeting quasars or from the BOSS ancillary science programs that has very strict
colour selections~(see Appendix B2 of \citealp{2013AJ....145...10D}). Hence, the
sample is biased towards hot and warm WDs~(typically
$T_{\mathrm{eff}}>14,000$\,K for DAs, $T_{\mathrm{eff}}>8,000$\,K for DBs; and a
minimum of $T_{\mathrm{eff}}=6,000$\,K). Thus, these catalogues are of little
use when it comes to the faint end of the WDLF which reveals the star formation
scenario of the Galaxy at early times. The use of reduced proper motion~(RPM) as
a proxy-absolute magnitude can separate WDs from the MS stars in an RPM diagram,
which resembles an HR diagram where the WDs are a few magnitudes fainter than
the MS stars. High speed digital imaging allows rapid scanning
of the sky at high cadence and to detect objects below the sky brightness, such
that the survey volume is greatly increased for the search of these faint
objects. This selection method has been proven to be efficient in identifying WD
candidates~(e.g.,~\citealp{1992MNRAS.255..521E}; \citealp{1999MNRAS.306..736K};
H06 and RH11). Although this technique gives more leverage to separate WDs from
MS stars, it is more difficult to treat completeness and
contaminations because of the introduction of an extra parameter -- proper
motion.

High quality proper motion requires a long maximum time baseline, large number
of epochs and high astrometric precision. A simplified proper motion uncertainty
relation can be approximated by $\sigma_{\mu} = \sqrt{2} \times \sigma_{x}
\times \frac{1}{\Delta t} \times \sqrt{\frac{12}{N}}$ where the $\sqrt{2}$ comes
from the symmetric contribution from the $\alpha$ and $\delta$ directions,
$\sigma_{x}$ is the astrometric precision, $\Delta t$ is the maximum epoch
difference, $N$ is the number of detections and the factor of $12$ comes from the
variance of a uniform distribution~\citep{2013ASPC..469..253H}. Most previous works, with the exception of a few~(e.g.,~\citealp{1999ASPC..165..413G}, M17 etc.), used
entirely or some photographic plate data in order to gain sufficient maximum
epoch difference, so the faint magnitude limit is roughly at the sky brightness,
$R\approx19.5$\,mag. This has significantly restricted the survey volume: in
H06, even though the photometry is given by the SDSS, the pairing criterion
limits the depth of their catalogue to the magnitude limits of the USNO-B1.0
survey; in RH11, the SuperCOSMOS Sky Survey was compiled by digitising several
generations of photographic plate surveys which has roughly the same photometric
limits in H06. Using the state-of-the-art Panoramic Survey Telescope and Rapid
Response System 1~(Pan--STARRS~1 or PS1, \citealt{2010SPIE.7733E..0EK}), with multi epoch data which has on
average $60$ epochs, proper motion objects were not limited to the ones that were
also detected in the past by photographic plates. This system can provide a
homogeneous selection of WD candidates.

This article is organised in the following structure. In Section~2, the properties of PS1 are described, and details how it  delivers a large sample of proper motion objects reaching the survey magnitude limits. The data selection by the derived properties  is described in Section~3. The technique for maximising the survey volume and the mathematical construction of the WDLF with the Voronoi method are detailed in Section 4. Section 5 presents the WDLFs of the solar neighbourhood and the halo. Section 6 compares the WDLFs with previous works. The final section finishes with a summary and a brief discussion.

\section{Selection Criteria - Survey Properties}
\label{sec:selection_criteria_survey}
The PS1 is a wide-field optical imager devoted to survey
operations~\citep{2010SPIE.7733E..0EK, 2016arXiv161205560C}. The telescope has a 1.8\,m diameter
primary mirror and is located on the peak of Haleakal$\bar{\mathrm{a}}$ on
Maui~\citep{2004SPIE.5489..667H}. The site and optics deliver a point spread
function~(PSF) with a full-width at half-maximum~(FWHM) of $\sim$$1\arcsec$ over a seven
square degree field of view. The focal plane of the telescope is equipped with
the Gigapixel Camera 1, an array of sixty $4,800\times4,800$ pixels orthogonal
transfer array~(OTA) CCDs~\citep{2009amos.confE..40T,2008SPIE.7014E..0DO}. Each
OTA CCD is further subdivided into an $8\times8$\,array of independently
addressable detector regions, which are individually read out by the camera
electronics through their own on-chip amplifier. Most of the PS1 observing time
is dedicated to two surveys: the 3$\upi$ Sterdian Survey~(3$\upi$ Survey), that covers the entire
sky north of declination $-30^{\circ}$, and the Medium-Deep Survey~(MDS), a
deeper, multi-epoch survey of $10$ fields, each of $\sim$$7$ square degrees in
size~\citep{2012AAS...22010704C}. Each survey is conducted in five broadband
filters, denoted g$_{\mathrm{P}1}$, r$_{\mathrm{P}1}$, i$_{\mathrm{P}1}$,
z$_{\mathrm{P}1}$ and y$_{\mathrm{P}1}$, that span over the range of
$400-1,000$\,nm. These filters are similar to those used in the SDSS, except the
g$_{\mathrm{P}1}$ filter extends $20$\,nm redward of g$_{\mathrm{SDSS}}$ while
the z$_{\mathrm{P}1}$ filter is cut off at $920$\,nm. The y$_{\mathrm{P}1}$
filter covers the region from $920$ to $1,030$\,nm where SDSS does not have an
equivalent one. These filters and their absolute calibration in the context of
PS1 are described in \citet{2012ApJ...750...99T}, \citet{2012ApJ...756..158S}
and \citet{2013ApJS..205...20M}. The PS1 images are processed by the PS1 Image
Processing Pipeline~(IPP; \citealp{2006amos.confE..50M, 2016arXiv161205240M}). This pipeline performs
automatic bias subtraction, flat fielding, astrometry, photometry,
and image stacking and differencing for every image taken by the system~\citep{2007ASPC..364..153M, 2008IAUS..248..553M, 2016arXiv161205245W, 2016arXiv161205244M, 2016arXiv161205242M}.
 
Each observation of the 3$\upi$ Survey visits a patch of sky two times with an interval of
15 minutes in between, which make a transit-time-interval\,(TTI)
pair~\citep{2012AAS...22010704C}. These observations are used primarily to
search for high proper-motion solar system objects~(asteroids and
Near-Earth-Objects). As part of the nightly processing these TTI pairs are
mutually subtracted and objects detected in the difference image are reported to
the Moving Object Pipeline Software. Each of the TTI pairs are taken at exactly
the same pointing and rotation angle so that the fill factor for searching for
asteroids is not compromised. However, the other TTI pairs are taken at 
different rotation angles and centre offsets such that a stack fills in
the gaps and masked regions of the focal plane. The g$_{\mathrm{P}1}$,
r$_{\mathrm{P}1}$ and i$_{\mathrm{P}1}$ bands are observed close to opposition
to enable asteroid discovery while the z$_{\mathrm{P}1}$ and y$_{\mathrm{P}1}$
bands are scheduled as far from opposition as feasible in order to enhance the
parallax factors of faint, low-mass objects in the solar neighbourhood. Each
year, the field is then observed a second time with the same filter for an
additional TTI pair of images, making four images of each part of the sky, in
each of the five PS1 filters, giving an average of $20$ images on $3\pi$
steradian of the sky per year. The positions given are corrected for differential chromatic refraction~(DCR). This Section describes all the selection criteria
based on the survey properties, where further selection requirements based on
the derived properties will be discussed in Section~\ref{sec:selection_criteria_derived}.

\subsection{Proper Motion}
Before the Gaia DR2~\citep{2018A&A...616A...1G, 2018A&A...616A...2L} became available most objects did
not have parallax measurements and WDs could only be identified efficiently with
proper motions. Therefore, on top of magnitude limits, a good knowledge of the 
proper motions and their associated uncertainties are needed to apply a completeness 
correction to a proper motion-limited sample. Beyond $\sim$$70$\,pc, the parallax solution from PS1 is the manifestation of 
amplified noise~\citep{2008IAUS..248..553M}. In particular, WDs are much fainter than stellar objects so they have even larger uncertainties at the same distance. The reliable distance estimation limit is even smaller.

There are large correlated errors between parallax and proper motion particularly when coverage in 
parallactic factor is low. These correlations are not propagated into the final catalogue products 
in PS1 PV2, so the proper motions from the 5-parameter solutions~(the pair of zero-point 
in the right ascension, $\alpha$, and declination, $\delta$, directions, the pair of proper 
motions and the parallax) have increased scatter over those that can be computed from 
the 4-parameter solutions using the epoch astrometry. Since the given set of astrometric 
solution is only good up to a few tens of parsecs, even for the study of nearby WDs, most 
of them lie outside the range where the parallax solutions are meaningful. Therefore, for 
our purposes we are required to compute our own set of 4-parameter solutions, for all sources with better than $1\sigma$ proper motion, where parallaxes 
are not solved for. The best fit solution is found by the method of least squares, when written 
in matrix form,
\begin{equation}
\underbrace{
\left(
\begin{array}{cccc}
\frac{1}{w_{0}} & 0 & \frac{t_{0}}{w_{0}} & 0 \\
0 & \frac{1}{w_{0}} & 0 & \frac{t_{0}}{w_{0}} \\
\cdot & \cdot & \cdot & \cdot \\
\cdot & \cdot & \cdot & \cdot \\
\frac{1}{w_{n}} & 0 & \frac{t_{n}}{w_{n}} & 0 \\
0 & \frac{1}{w_{n}} & 0 & \frac{t_{n}}{w_{n}}
\end{array}
\right)
}_\mathrm{\bf A}
\left(
\begin{array}{c}
\xi_{\mathrm{ZP}}\\
\eta_{\mathrm{ZP}}\\
\mu_{\xi}\\
\mu_{\eta}
\end{array}
\right)
=
\left(
\begin{array}{c}
\frac{\Delta\xi_{0}}{w_{0}}\\
\frac{\Delta\eta_{0}}{w_{0}}\\
\cdot\\
\cdot\\
\frac{\Delta\xi_{n}}{w_{n}}\\
\frac{\Delta\eta_{n}}{w_{n}}
\end{array}
\right)
\end{equation}
with
\begin{equation*}
w_{i} = \sqrt{ \Delta m_{i}^{2} + 0.015^{2} }
\end{equation*}
where $w_{i}$ is the weight, $t_{i}$ is the epoch of the measurement, $\xi$ and $\eta$ are the local plane coordinates in the direction of the right ascension and declination, $\Delta\xi_{i}$ is the 
offset of the $\xi_{i}$ from the mean position, $\Delta\eta_{i}$ is that for 
$\eta_{i}$, $\sigma_i$ is the astrometric precision, $0.015$ is the noise floor of the 
PV2 photometry and $\Delta m_{i}$ is the photometric uncertainty. The solutions are the 
$\xi$, $\eta$, $\mu_{\xi}$ and $\mu_{\eta}$ in the middle bracket. The associated 
uncertainties are diagonal terms of the dot product of the transpose of first matrix {\bf A} with itself,
\begin{equation}
\left(
\begin{array}{c}
\sigma_{\xi}^2 \\
\sigma_{\eta}^2 \\
\sigma_{\mu_{\xi}}^2 \\
\sigma_{\mu_{\eta}}^2
\end{array}
\right)
=
\mathrm{diag}\left[\left({\bf A}^{\mathrm{T}} {\bf A}\right)^{-1}\right].
\end{equation}
The re-computation of
proper motion is performed on all objects with signal-to-noise~(S/N) ratio greater than unity in
the given proper motions.

\subsection{Reduced Proper Motion (RPM)}
\begin{figure}
\includegraphics[width=84mm]{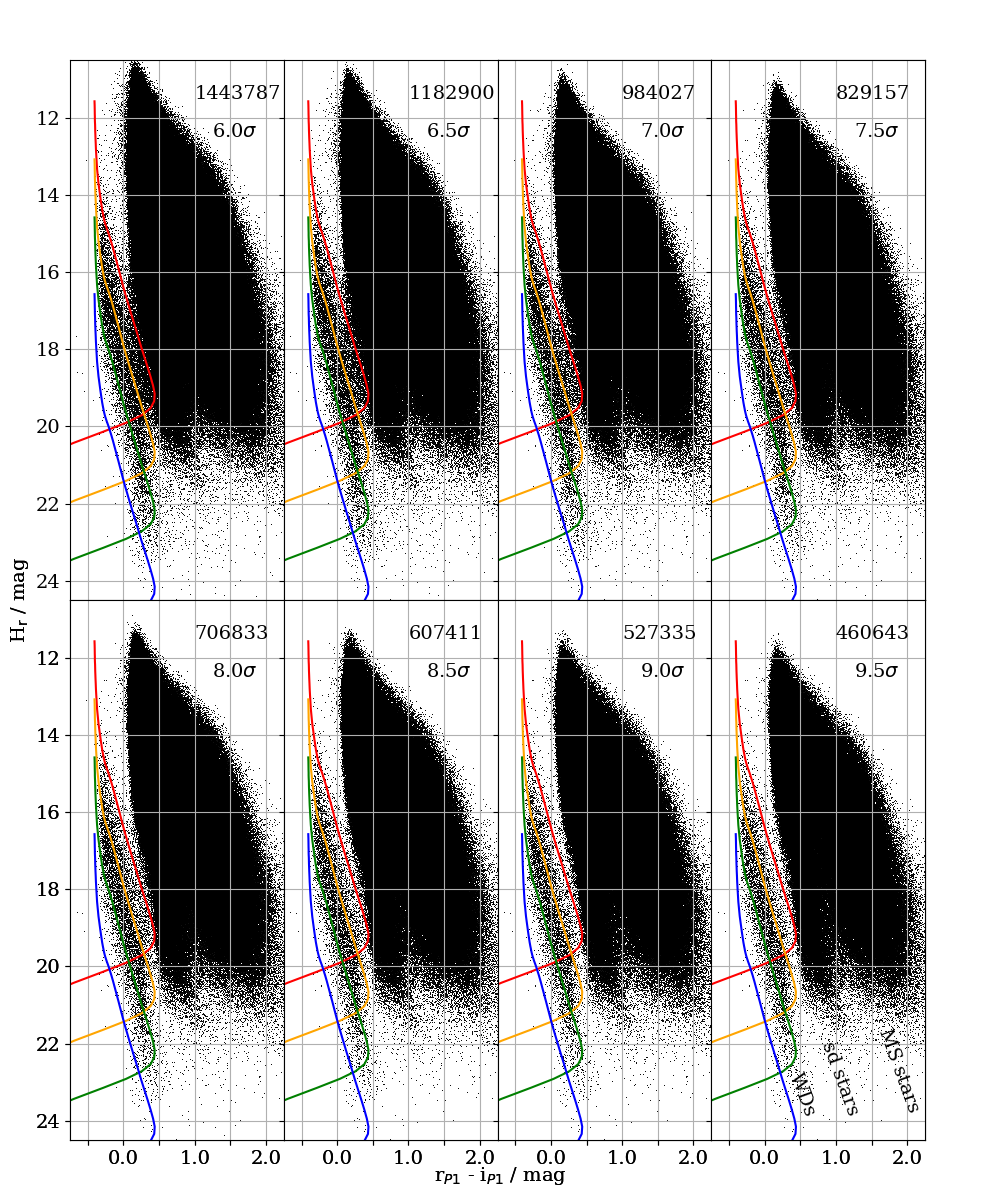}
\caption{RPM diagrams with $\mu > 5.0$ to $9.5\sigma$ from top left to bottom right. The numbers above the significance shows the total number of objects in the scattered plots.
The red, yellow, green and blue lines are the cooling sequence for DA WDs with tangential velocity 
at $20, 40, 80$ and $200\kmps$. Most contaminants appear as vertical scatter with neutral 
colour~($0.0<$r$_{\mathrm{P}1}-$i$_{\mathrm{P}1}<0.5$).}
\label{fig:rpmd_compared}
\end{figure}

There exists a correlation between proper motions and distance of nearby
objects, since closer objects are more likely to show large proper motions. RPM,
$H$, combines the proper motion with apparent magnitude to provide a crude
estimate of the absolute magnitude. Thus, the RPM equation has a close
resemblance to the absolute-apparent magnitude relation,
\begin{align}
H_{m} &= m + 5\log \mu + 5 \\
      &= M + 5\log v_{\mathrm{tan}} - 3.3791
\label{eq:rpm}
\end{align}
where $\mu$ is the proper motion in arcseconds per year, $m$ is the apparent
magnitude, $M$ is the absolute magnitude and $v_{\mathrm{tan}}$ is the
tangential velocity in kilometers per second. The RPM of WDs are a few
magnitudes fainter than MS dwarf and subdwarf stars of the same colour.
Therefore, the WD locus is separated from other objects in the RPM diagram. This has been proved
to be an efficient way to obtain a clean sample~(e.g.,~H06 and RH11). In this work, we
use the r$_{\mathrm{P}1}$ to calculate the RPM, which is denoted by $H_{r}$;
see Fig.~\ref{fig:rpmd_compared} for RPM diagrams at different 
levels of proper motion significance.

\subsection{Lower Proper Motion Limits}
\label{sec:lower_pm_limit}
In order to select a clean sample of proper motion objects, we require our
samples to have high S/N ratio in the proper motions. This excludes
most of the non-moving objects from our catalogue and limits scatter in the RPM
diagram. The total proper motion uncertainty of an individual object is
given by
\begin{equation}
\sigma_{\mu} = \sqrt{\left(\frac{\mu_{\alpha}\cos(\delta)}{\mu}\right)^{2}\sigma_{\mu_{\alpha}\cos(\delta)}^{2}+\left(\frac{\mu_{\delta}}{\mu}\right)^{2}\sigma_{\mu_{\delta}}^{2}} ,
\end{equation}
where $\mu$ is the total proper motion.

When $\sigma_{\mu}$ is plotted against r$_{\mathrm{P}1}$ there is
significant scatter at a given magnitude. However, a well defined relation
between the proper motion uncertainty and magnitude is needed for volume
integration and completeness corrections~(Section \ref{sec:Vmod}). See Section~\ref{sec:volume_voronoi} for how the individual lower proper motion limits can be applied to a sample from a non-uniform survey.

\subsection{Upper Proper Motion Limit}
The upper proper motion limit is determined by PS1 PV2 matching radius,
matching algorithm and its efficiency.

\subsubsection*{Matching Radius}
The search radius for cross-matching between different epochs of PS1 data was
$1\arcsec\,yr^{-1}$. Although each part of the sky was imaged twelve times per year
on average, some parts of the sky were limited by seasonal observability and
weather. At low declinations, the sky could only be observed in a window of a
few months every year so the maximum proper motion an object can carry is
limited to roughly the size of the search radius per year, which is
$1\arcsec\,yr^{-1}$.

\subsubsection*{Matching Algorithm}
In PV2 high proper motion objects moving by more than $1\arcsec$
throughout the survey period would be detected as 2 or more separate
objects. The IPP solves for the 5-parameter astrometric solutions that include
parallax. In order to break the degeneracy in the parallax and proper motion in
the astrometric solution, a minimum epoch difference of 1.5 years is required.
For objects that move faster than $0.66\arcsec\,yr^{-1}$, they would have moved
outside the matching radius after 1.5 years. Otherwise, these objects would have
either erroneously large proper motion with small parallax or vice versa.
Although it is possible to ``stitch'' the multiple parts back together and
recalculate the proper motions with the maximal use of data, this creates a
completeness problem to the faint high proper motion objects. When objects close
to the detection limits can only be observed under the best observing
conditions, there are not enough epochs to solve for the astrometric solution
when they are split into parts. For example, if an object has 10 evenly
distributed measurements that are catalogued as ``2 sources'' each with 5
measurements, the individual uncertainty would become
$2\times\sqrt{2}\approx2.28$ times larger than that is solved as a single object
where the $2$ comes from the ratio of the maximum epoch difference and
$\sqrt{2}$ comes from the ratio of the number of epochs.

\subsubsection*{Matching Efficiency}
\begin{figure}
\includegraphics[width=84mm]{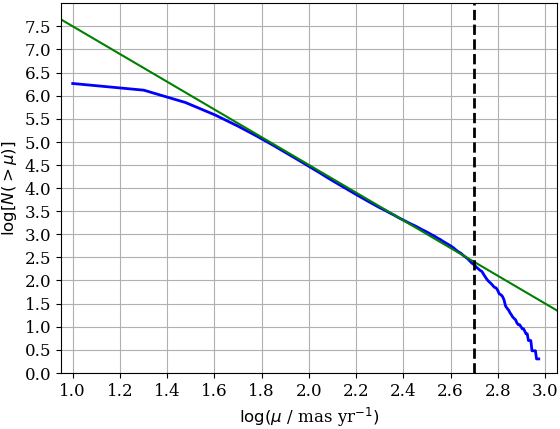}
\caption[Number of object with proper motion larger than the given proper motion against proper motion]
{The logarithm of the number of object with proper motion larger than the given proper motion is plotted 
against proper motion~(blue). Due to the small distances the high proper motion objects are at, the relation 
has a gradient of $-3$ as shown by the green line. Above $10^{2.7}=501\,\mathrm{mas}\,\mathrm{yr}^{-1}$~(dashed line), the
 underestimation of the number of objects implies an incomplete matching of those high proper motion objects}
\label{fig:upper_pm_limit}
\end{figure}

The high proper motion population is in the immediate solar neighbourhood so the
number density is uniform at this limit. Through using proper motion as a
proxy-parallax~(like that in reduced proper motion), the number density follows
\begin{equation}
\log{N} \propto -3 \log(\mu)
\end{equation}
for a complete sample. In the 3$\upi$ Survey, the gradient deviates from $-2.7$ at 
$0.501\arcsec\,yr^{-1}$~(See Fig. \ref{fig:upper_pm_limit}).

Combining the three cases, the matching efficiency gives the tightest limit among
all, so the global upper proper motion limit in this study is set at
$0.501\arcsec\,yr^{-1}$.

\subsection{Faint Magnitude Limit}
\begin{figure}
\includegraphics[width=84mm]{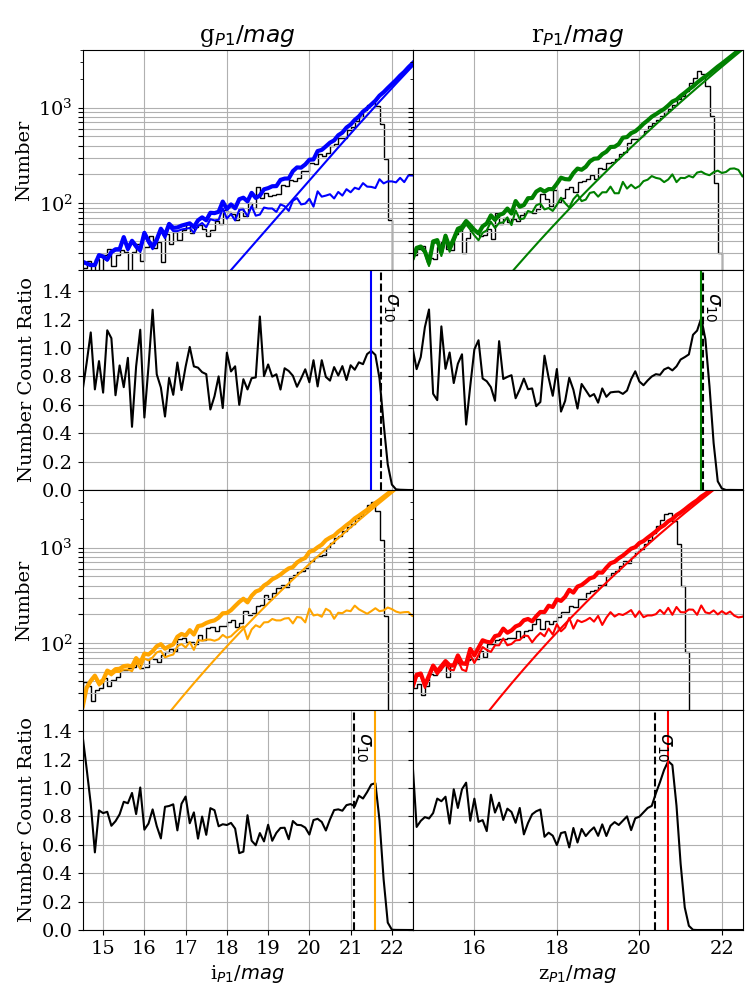}
\caption[Observed star and galaxy counts compared with model counts]
{An example star and galaxy counts in g$_{\mathrm{P}1}$~(purple), r$_{\mathrm{P}1}$~(blue), 
i$_{\mathrm{P}1}$~(green) and z$_{\mathrm{P}1}$~(yellow) filters in the direction $(\alpha,\delta)=(0.0,-13.248015)$. The odd rows (from top) show 
the star and galaxy counts with thin colour lines and the combined star and galaxy 
counts in thick colour line. The black lines are the observed number counts. The even 
rows show the ratios between the model and observation, the dashed lines show the 
10$\sigma$ photometric limits and the colour lines mark the completeness limit~(see 
Section~\ref{sec:completeness}).}
\label{fig:star_counts}
\end{figure}

In order to find the faint magnitude limits at which data are complete, the
object counts were compared against synthetic star and galaxy counts in
g$_{\mathrm{P}1}$, r$_{\mathrm{P}1}$, i$_{\mathrm{P}1}$ and z$_{\mathrm{P}1}$
filters in 15 fields at high galactic latitudes to avoid interstellar extinction complicating 
this analysis. We chose a field of view of $\sim$$3.4$ square
degrees~(a HEALPix pixel size with $\mathrm{N}_{\mathrm{side}} = 32$, see later), a
size that is large enough for sufficient star counts and to smooth out 
inhomogeneity of galaxies while at the same time small enough
to limit variations in data quality across the field~(see Fig.
\ref{fig:star_counts}). Each of the filter is treated independently in this exercise where each source is required to be detected at least 3 times with a magnitude uncertainty less than $0.2\magnitude$.

\subsection*{Stars}
Differential star counts along the line of sight to each field were obtained
using the Besan\c{c}on Galaxy model~(\citealp{2003A&A...409..523R},
\citeyear{2004A&A...416..157R}). This employs a population synthesis approach to
produce a self-consistent model of the Galactic stellar populations, which can
be ``observed'' to obtain the theoretical star counts. It is a useful tool to
test various Galactic structure and formation scenarios although we have adopted
all the default input physical parameters except the latest spectral type is DA9
instead of the default DA5. There are only two photometric systems available,
the Johnson-Cousins and the CFHTLS-Megacam systems. Since there is only a small
difference between the PS1 and Megacam, the g', r', i' and z' are used to
approximate the g$_{\mathrm{P}1}$, r$_{\mathrm{P}1}$, i$_{\mathrm{P}1}$ and
z$_{\mathrm{P}1}$ in this work. The faint magnitude limits of the model are set
at $25\magnitude$ to guarantee that the model is always complete as compared to
the data.

\subsection*{Galaxies}
Fainter than $\sim$$19 \magnitude$, galaxies become
unresolved~(i.e.,~point-like) and have photometric parameters that overlap with
stars. Therefore, it is necessary to include galaxies in the synthetic number
counts. Galaxy counts to faint magnitudes have been determined in many
independent studies. The Durham Cosmology Group has combined their own results
(see e.g.,~\citealp{1991MNRAS.249..481J}; \citealp{1991MNRAS.249..498M}) with
many other authors. These are available online along with transformations to
different photometric
bands\footnote{http://astro.dur.ac.uk/$\sim$nm/pubhtml/counts/counts.html}. They
are provided in terms of log-number counts per square degree per half-magnitude
unless specified otherwise. A cubic spline was fitted over all available
observations to obtain the galaxy counts as functions of magnitude in each band.

\begin{figure}
\includegraphics[width=84mm]{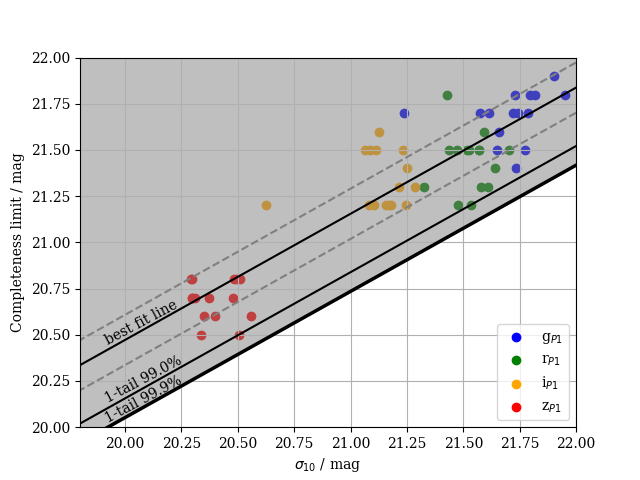}
\caption[The completeness limits against $10\sigma$ point source detection limits]
{The completeness limits are plotted against the $10\sigma$ point source detection limits. 
The solid line is the best fit linear relation, dashed line is the vertical offset of the 
best fit solution that has covered $68.2\%$ of all data~(i.e.,~$1\sigma$ in each direction).}
\label{fig:completeness_relation}
\end{figure}

PS1 has a very complex variation in the data quality as a function of position.
If the small/medium scale variations in the survey depth are not considered, the
survey volume would be limited to the shallowest parts of the sky, which would
be more than a magnitude brighter than the deepest parts. In order to take into
account these small scale effects, a linear relationship between the
completeness magnitude and the detection depth map~\citep{2014MNRAS.437..748F},
$D(\alpha,\delta)$, was found empirically, see
Fig.~\ref{fig:completeness_relation}. Since the given depth maps are the
$10\sigma$ detection limit in a fiducial $3\arcsec$ aperture we converted to
FWHM magnitude by accounting for the flux included in the PSF, so the limiting
magnitude was corrected with a linear transformation $D' = D -
2.5\times\log(2)$. The characteristics in the y$_{\mathrm{P}1}$ band were
assumed similar to the other filters.

\subsection{Survey Depth}
\label{sec:completeness}
The $3.4$ square degrees field-of-view of PS1 is not small compared to the size
of inhomogeneities in survey quality so there is some scatter in the
completeness-depth relation. In order to account for these variations, instead
of choosing the best fit straight line~($C$, where $C = 0.6826 \times \sigma_{10} + 6.8197$), which has half the data points above the
line and the other half below it, a straight line that would have covered 99.9\%
of all data was used, this corresponds to $3.090\,\sigma_{\mathrm{measured}}$. This threshold means that $99.9\%$ of the time the HEALPix pixel is complete. The scatter of these points was measured from the median
absolute deviation~(MAD) to minimize the effect from outliers, where
$\sigma_{\mathrm{measured}} = 1.48\times \mathrm{MAD}$. The completeness limit is
\begin{equation}
C' = C - 3.090\times\sigma
\end{equation}
where $\sigma$ is measured to be $0.1359$.

By applying this relation to the photometric depth maps, the completeness maps
in the five PS1 filters were produced. The resolution at which these maps
were applied was degraded to
$\mathrm{N}_{\mathrm{side}}=16$ to match the resolution of the tangential
velocity completeness correction.

\begin{figure}
\includegraphics[width=82mm]{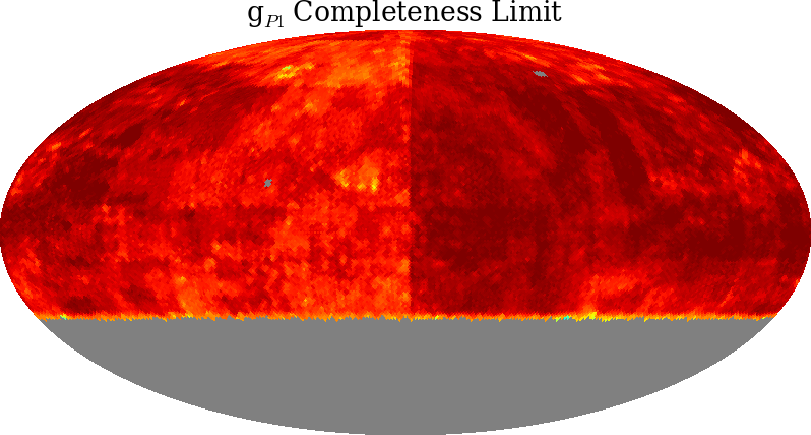}
\includegraphics[width=82mm]{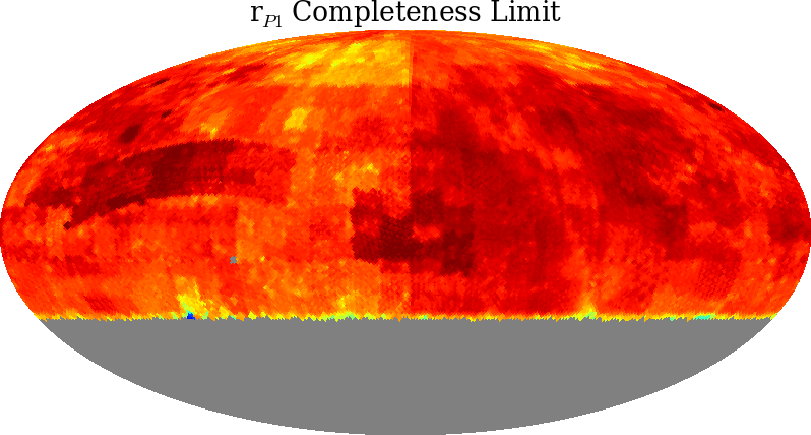}
\includegraphics[width=82mm]{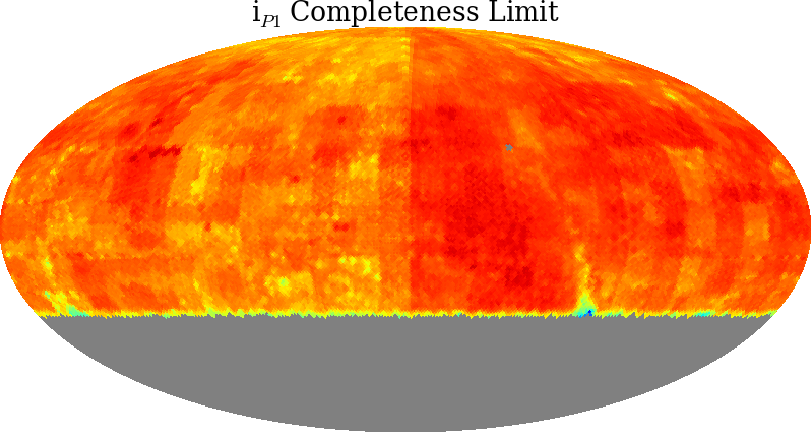}
\includegraphics[width=82mm]{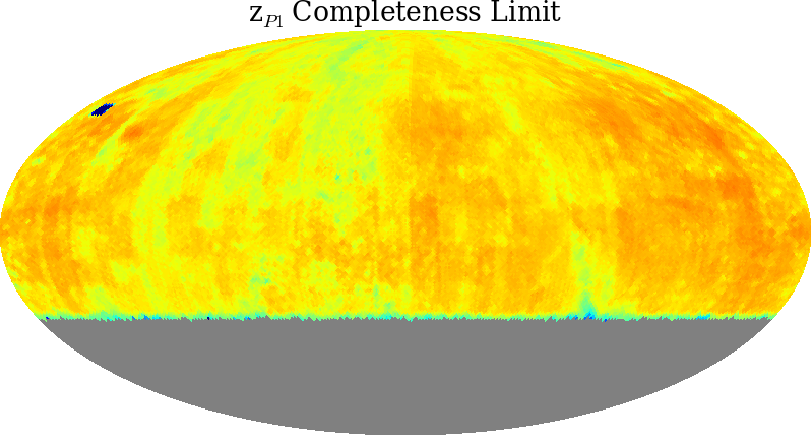}
\includegraphics[width=82mm]{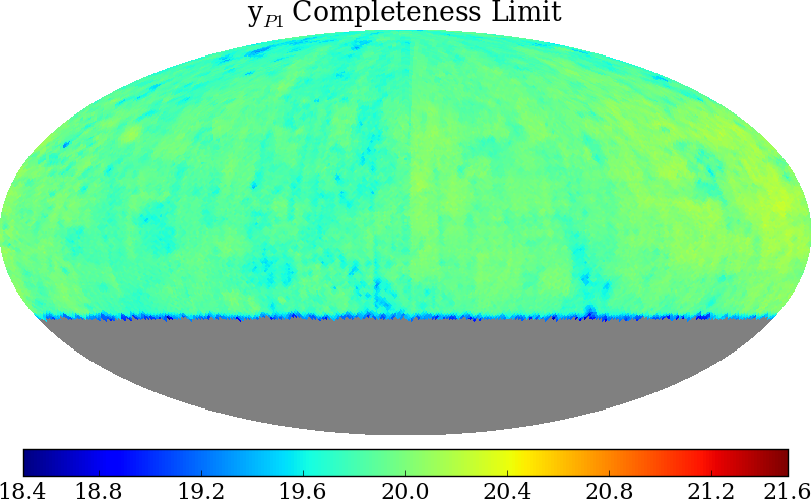}\\
\caption[Completeness map in g$_{\mathrm{P1}}$, r$_{\mathrm{P1}}$, i$_{\mathrm{P1}}$, 
z$_{\mathrm{P1}}$ and y$_{\mathrm{P1}}$ filters]
{Completeness map in g$_{\mathrm{P1}}$, r$_{\mathrm{P1}}$, i$_{\mathrm{P1}}$, 
z$_{\mathrm{P1}}$ and y$_{\mathrm{P1}}$ filters respectively, all magnitudes are with the same colour scale.}
\label{fig:depth_map}
\end{figure}

\subsection{Bright Magnitude Limit}
Brighter than $15\magnitude$, there is an astrometric bias that is colloquially known as
the `Koppenhofer effect' amongst the PS1 Science Consortium. The essence of the effect was that a large charge packet could be drawn prematurely over an intervening negative
serial phase into the summing well, and this leakage was
proportionately worse for brighter stars. The brighter
the star, the more the charge packet was pushed ahead. The amplitude of the effect was at most $0.25''$, corresponding to a shift of about one pixel. Roughly a quarter of the data were affected before the problem was corrected~\citep{2016arXiv161205242M}. Since there are few
WDs brighter than $15\magnitude$, we choose this as our bright limit to minimise the effect.

\subsection{Object Morphology}
All objects marked as good were selected~(i.e.,~not flagged as
extended, rock, ghost, trail, bleed, cosmic ray or asteroid).

\subsubsection*{Star-Galaxy Separation}
The 3$\upi$ Survey catalogue has a star-galaxy separator~(SGS) entry for every object. The
typical photometric limits are $\sim$$21\magnitude$ in the optical and a typical
FWHM of $1.2\arcsec$ so at the faint end of the survey, the realiability of the SGS
is limited to the observing conditions. Therefore, we compared the SGS with the object classifier from the
Canada-France-Hawai'i Telescope Lensing Survey~(CFHTLenS,
\citealp{2012MNRAS.427..146H}; \citealp{2012AAS...21913009E} and
\citealp{2012MNRAS.421.2355H}) employing codes \texttt{CLASS\_STAR}, \texttt{star\_flag}
and \texttt{FITCLASS}. CFHTLenS is a $154$ square degrees multi-colour optical survey with
the Megacam u*, g', r', i' and z' filters incorporating all data collected in
the five-year period on the CFHT Legacy Survey, which was optimised for weak
lensing analysis. The deep photometry in the i'-band was always taken in
sub-arcsecond seeing conditions. Both \texttt{star\_flag} and \texttt{FITCLASS} were optimised for
galaxy selection, so the \texttt{CLASS\_STAR} provided by \textsc{SExtractor} was used in this
analysis. Considering the superior quality in both photometry and observing
conditions of CFHTLS, at the limit of i$_{\mathrm{P1}}\sim$$21\magnitude$, we
assumed that \texttt{CLASS\_STAR} was completely reliable. The pairing criteria of the
two catalogues were $2''$ matching radius and $5\sigma$ proper
motions.

\begin{figure}
\includegraphics[width=84mm]{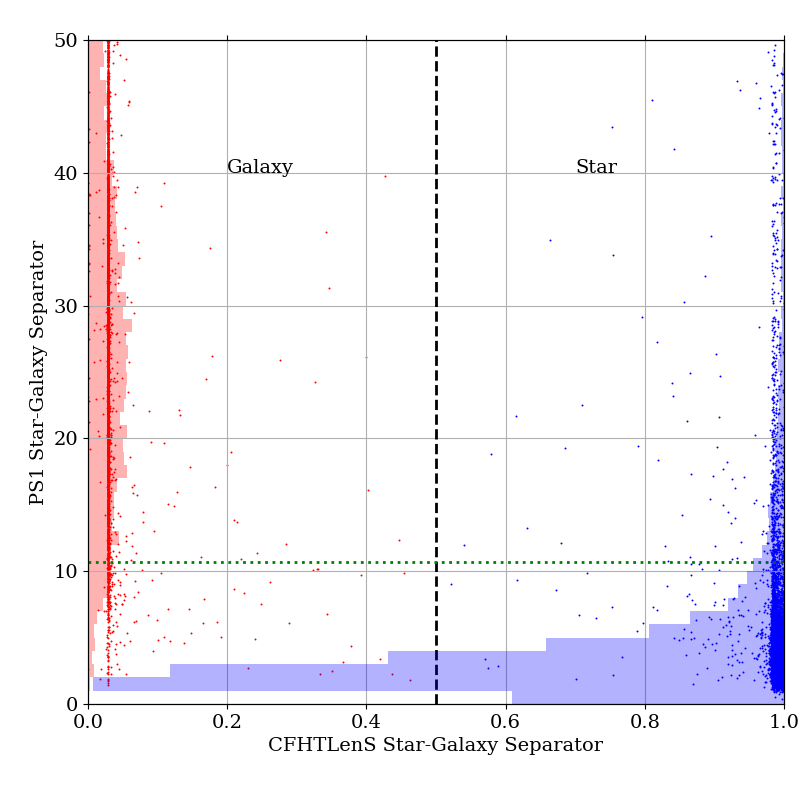}
\caption[CFHTLenS star galaxy separator, \texttt{CLASS\_STAR}, plotted against the PS1 star-galaxy separator]
{PS1 SGS is plotted against the CFHTLenS star galaxy separator, \texttt{CLASS\_STAR}. When the \texttt{CLASS\_STAR} 
is larger than $0.5$, it is considered as a star; otherwise, a galaxy. The green dotted line indicates the PS1 SGS limit. The semi-transparent histograms are the respective number counts as functions of PS1 SGS, the full range of the x-axis corresponds to a number count of $2,500$.}
\label{fig:sgs_compared}
\end{figure}

In Fig.~\ref{fig:sgs_compared}, the PS1 SGS is plotted against \texttt{CLASS\_STAR}. We
defined an object as a star when \texttt{CLASS\_STAR} $> 0.5$ or as a contaminant
otherwise. The green dotted line indicates the PS1 SGS limit at $10.728$ which keeps the sample at $90.0\%$ complete and with a galaxy contamination rate of $3.9\%$.

\section{Selection Criteria - Derived Properties}
\label{sec:selection_criteria_derived}
The construction of a WDLF depends on the distance, luminosity and atmosphere
type of the WDs, as well as the physical properties of the host population.
Since most detected WDs lie within a few hundred parsecs from the Sun, the
radial scale-lengths, which are of the order of kiloparsecs, of all Galactic
components were not considered in this work. Interstellar reddening was
corrected with the use of a three dimensional dust map when solving for the photometric
parallax.

\subsection{WD Atmosphere Type}
On the theoretical front, WD atmospheres have been studied in detail. In recent
years, with the abundant spectroscopic data available from SDSS, there were
significant improvements in the understanding in the atmospheres. In addition to
the conventional DA~($1,500$\,K $< T_{\mathrm{eff}}<120,000$\,K) models,
synthetic photometry is available for 9 different hydrogen-helium mass ratios in
the range $2,000$\,K
$<T_{\mathrm{eff}}<12,000$\,K~(\citealp{2006AJ....132.1221H};
\citealp{2006ApJ...651L.137K}; \citealp{2011ApJ...730..128T}; and
\citealp{2011ApJ...737...28B}\footnote{http://www.astro.umontreal.ca/$\sim$bergeron/CoolingModels}).
We choose the most helium rich model with $\log(\frac{\mathcal{M}_{\mathrm{He}}}{\mathcal{M}_{\mathrm{H}}} = 8.0)$ to be our DB model. All models
were provided in the PS1 filters by Dr.~Pierre Bergeron~(private communication). The
cooling tracks of different chemical compositions are very similar above
$T_{\mathrm{eff}}\sim10,000$\,K~(i.e.,~$\mathrm{M}_{\mathrm{bol}} < 12.0$).

\subsection{Interstellar Reddening}
A three dimensional map of interstellar dust reddening was produced using 800
million stars with PS1 photometry of which $200$ million also have 2MASS
photometry~\citep{2015ApJ...810...25G}. Although there is a health warning that
the reddening is ``{\it best determined by using the representative samples,
rather than the best-fit relation}'', with $\sim$$20,000$ spectroscopically confirmed WDs over the whole
sky, most of which reside in the SDSS footprint, the only way to deredden our samples was to use the given best-fit
solution. In order to convert the reddening values of $E(B-V)$ to extinction in
the PS1 photometric systems, the values on Table 6 of \citet{2011ApJ...737..103S} were
used. We adopted the values from the column $R_{v} = 3.1$ for this work~(see
Table \ref{table:reddening}).

\begin{table}
\caption{$A_{x}/E(B-V)_{SFD}$ in different passbands x, evaluated according to an 
\citet{1999PASP..111...63F} reddening law with $R_{v} = 3.1$ using a 7,000 K source spectrum. The subscript SFD refers to \citet{1998ApJ...500..525S}.}
\begin{tabular}{cccccc}
Passband~(x) & g & r & i & z & y\\
\hline
\hline
$A_{x}/E(B-V)_{SFD}$       & 3.172 & 2.271 & 1.682 & 1.322 & 1.087\\
\hline
\end{tabular}
\label{table:reddening}
\end{table}
The reddening information along the line of sight was given between distance
modulus $4.0$ and $19.0$ in $0.5$ intervals. Each line of sight was interpolated with
a cubic spline between the given points in order to compute the reddening at 
arbitrary distance.

\subsection{Photometric Parallax}
\begin{figure}
\includegraphics[width=84mm]{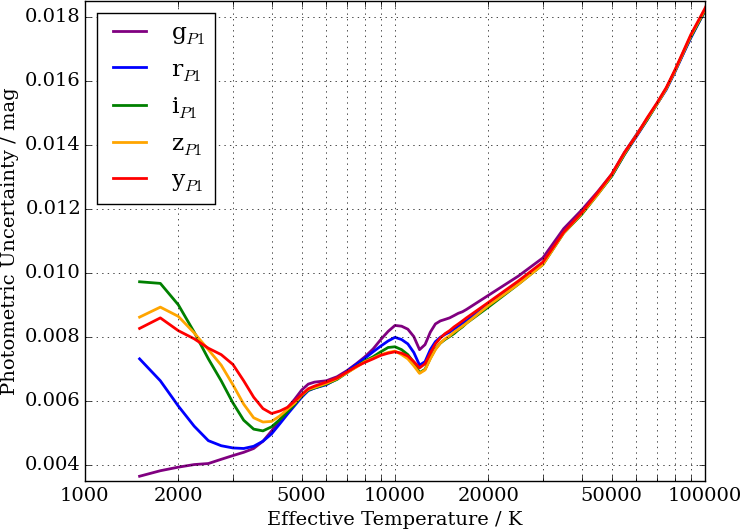}
\caption[The standard deviations in magnitude in each filter when a population 
is assumed to have fixed surface gravity of $\log(g)=8.0$ for DA WDs]
{The standard deviations in magnitude in each filter when a population following 
the distribution described by equation \ref{eq:logg} is assumed to have fixed 
surface gravity of $\log(g)=8.0$ for DA WDs.}
\label{fig:fixed_logg}
\end{figure}

The surface gravities of WDs are narrowly distributed at about
\begin{equation}
\label{eq:logg}
\left<\log\,g\right>=7.937\pm0.012
\end{equation}
with SDSS DR10~\citep{2015MNRAS.446.4078K}. Thus, by assuming a constant surface
gravity at $\log(g)=8.0$, the distance and temperature of an object can be
determined simultaneously. However, in doing so, extra scatter is introduced to
the solution statistics. The goodness-of-fit $\chi^{2}_{\nu}$ would not be at
$\sim$$1$. Therefore, a simple Monte Carlo method was used to produce a table
of WDs following the distribution of equation \ref{eq:logg}. The standard
deviations in magnitudes in each of the filters were found as a function of
temperature for each of the DA and DB models~($\sigma_{log(g)}$). With these
relations, it was possible to propagate the uncertainties arisen from adopting
constant surface gravity into the final photometric parallax solutions~(Fig.~\ref{fig:fixed_logg}).
This is also important in up/down-weighting different filters in the fitting procedure where
the WD SEDs vary most significantly in the bluer filters.

The best-fit solutions with the DA and DB atmospheres were found by sampling
the distance-temperature space with a Markov-Chain
Monte-Carlo~(MCMC) method
\textsc{emcee}\footnote{http://dan.iel.fm/emcee}~\citep{2013PASP..125..306F}.
In both cases, we used $20$ walkers of length $20,000$ with a burn-in phase of $1,000$ steps.
There are some degeneracies in the solution, the most notable one being
between a cool WD at small distance and a hot WD at large distance because
interstellar reddening alters the shape of the model spectral energy
distribution~(SED). A simple minimisation technique~(e.g.,~with Nelder-Mead method) sometimes cannot guarantee a global minimal in some cases.

When interstellar reddening is included in the calculation, the likelihood function to be maximised is
\begin{equation}
\begin{aligned}
\sum_{i} &\left\{ \frac{ \left[ m_{i} - \mu_{D} - m_{\mathrm{model},i}(T_{\mathrm{eff}}) - A_{i}(D) \right]^{2} }{ \sigma_{i}^{2} + \sigma_{log(g),i}^{2} } + \right.\\
&\hspace*{5mm} \log\left[2\pi(\sigma_{i}^{2} + \sigma_{log(g),i}(T_{\mathrm{eff}})^{2})\right] \Bigg\}
\end{aligned}
\end{equation}
where $m_{i}$ is the magnitude filter $i$, $\mu_{D} = 5 \log(D) - 5$ is the
distance modulus with subscript $D$ to distinguish it from the symbol for proper
motion, $m_{\mathrm{model},i}(T_{\mathrm{eff}})$ is the magnitude of a given
model which depends only on the effective temperature and $A_{i}(D)$ is the
total extinction at distance $D$. In the case of the mixed atmosphere models,
the model magnitude becomes
$m_{\mathrm{model},i}(T_{\mathrm{eff}})$.

\section{Survey Volume Maximisation}

There are various statistical methods to arrive at a luminosity function. The
most commonly used estimator in WD studies is the maximum volume density
estimator \citep{1968ApJ...151..393S}. Its relatively straightforward approach has attributed to its popularity. This method was developed to combine
several independent surveys \citep{1980ApJ...235..694A} and to correct for
scaleheight effects~\citep{1989MNRAS.238..709S, 1993ApJ...414..254T}.
\citet{2006MNRAS.369.1654G} has shown that it is superior to the Cho\l{}oniewski
and the Stepwise Maximum Likelihood method at the faint end of the WDLF,
provided that the sample is sufficiently large, with more than $300$ objects. Like
many previous works, when objects are both photometric and proper motion
limited, extra caution is needed in order not to introduce bias. Simple,
but sufficient at the time, assumptions were made to cope with such cases
\citep{1975ApJ...202...22S}. However, it was shown in \citealp{2015MNRAS.450.4098L}
~(hereafter LRH15) that the estimator
underestimates the density of the intrinsically faint objects, and the modified maximum volume
should be used where the discovery fraction is inseparable from the volume
integrand. \citet[][hereafter L17]{2017MNRAS.469.1026L} further extended the LRH15 method to conduct object selection based on individual proper motion uncertainties before which a global uncertainty has to be used in order to perform completeness correction.

The maximum volume density estimator~\citep{1968ApJ...151..393S} tests the observability of a source by finding the maximum volume in which it can be observed by a survey~(e.g.,~at a different part of the sky at a different distance). It is proven to be unbiased~\citep{1976ApJ...207..700F} and can easily combine multiple surveys~\citep{1980ApJ...235..694A}. In a sample of proper motion sources, we need to consider both the photometric and astrometric properties~(see LHR15 for details). The number density is found by summing the number of sources weighted by the inverse of the maximum volumes. For surveys with small variations in quality from field to field and from epoch to epoch, or with small survey footprint areas, the survey limits can be defined easily. However, in modern surveys, the variations are not small; this is especially true for ground-based observations. Therefore, properties have to be found locally to analyse the data most accurately. Through the use of Voronoi tessellation, sources can be partitioned into individual 2D cells within which we assume the sky properties are defined by the governing source. Each of these cells has a different area depending on the projected density of the population.

HEALPix is the acronym for \underline{H}ierarchical \underline{E}qual
\underline{A}rea iso\underline{L}atitude \underline{Pix}elization of a
sphere~\citep{2005ApJ...622..759G}. This pixelisation routine produces a
subdivision of a spherical surface in which all pixels at the same level
in the hierarchy cover the same surface
area. All pixel centres are placed on rings of constant latitude, and are
equidistant in azimuth~(on each ring). However, the pixels are not regular in
shape. A HEALPix map has $\mathrm{N}_{\mathrm{pix}} = 12 \mathrm{N}_{\mathrm{side}}^{2}$
pixels each with the same area $\Omega =
\pi / 3\mathrm{N}_{\mathrm{side}}^{2}$, where $\mathrm{N}_{\mathrm{side}}$
is the square root of the number of division of the base pixel and it can be any value with a base of 2~(i.e.,~$2^{x}$
for any positive integer $x$). This pixelisation routine is used in computing the tangential velocity completeness correction.

In the rest of the article, {\it cell} will be used to denote Voronoi cell and {\it h-pixel} for HEALPix pixel.

\subsection{Tangential Velocity Completeness Correction}
In order to clean up the sample of proper motion objects, a lower tangential
velocity limit was applied to remove spurious sources~(low-velocity WDs have similar
RPMs to those of high velocity subdwarfs from the Galactic halo). For example,
$20, 30$ or $40\kmps$ are typical choices to obtain clean samples of the disc
populations, the precise choice of the value depends on the data quality; and
$160, 200$ or $240\kmps$ are used to obtain stellar halo objects~(H06, RH11, M17).
However, this process removes genuine objects from the sample. With some
knowledge of the kinematics of the solar neighbourhood, it is possible to model
the fractions of objects that are removed in any line of sight. A resolution of
$\mathrm{N}_{\mathrm{side}}=16$ was used to pixelise the sky into $3,072$ {\it h-pixels}
in order to account for the variation in the projected kinematics across the sky.

The problem of incompleteness as a result of kinematic selection bias was identified by \citet{1986ApJ...308..347B}. A
Monte--Carlo\,(MC) simulation was used to correct for such incompleteness by
comparing with star counts. This correction, known as the discovery fraction,
$\chi$, was then applied by H06. Instead of using a simulation,
\citet{2003MNRAS.344..583D} arrived at the discovery fractions by integrating
over the Schwarzschild distribution functions to give the tangential velocity
distribution, $\mathrm{P}(v_{\mathrm{T}}, \alpha, \delta)$. This was done by
projecting the velocity ellipsoid of the Galactic populations on to the tangent
plane of observation, correcting for the mean motion relative to the Sun, and
marginalising over the position angle to obtain the distribution in tangential
velocity~(see \citealp{1983veas.book.....M}). The values adopted for the mean
reflex motions and velocity dispersion tensors are given in Table
\ref{table:galactic_structure}. These are obtained from the
\citet{2009AJ....137..266F} study of SDSS M dwarfs, with values taken from their
0--100\,pc bin that is least affected by the problems associated with the
deprojection of proper motions away from the plane \citep{2009MNRAS.400L.103M}.
RH11 further generalised the technique to cope with an all sky survey as opposed
to the individual fields of view employed in earlier works. However, there are
some discrepancies between the parameter space in which the volume and the
discovery fractions were integrated over in all these cases. In order to
generalise over a proper motion limited sample properly, the effects of the
tangential velocity limits and the proper motion limits have to be considered
simultaneously at each distance interval. The discovery fraction at a given distance,
$\chi(\alpha_h,\delta_h,r)$, can be found from the normalised cumulative
distribution function of the WD tangential velocity,
\begin{equation}
\label{eq:discovery_fraction}
\chi(\alpha_h,\delta_h,r) = \int_{a(r)}^{b(r)} \mathrm{P}(v_{\mathrm{T}},\alpha_h,\delta_h) \mathrm{d}v_{\mathrm{T}}
\end{equation}
where
\begin{equation}
a(r) = \mathrm{max}(v_{\mathrm{min}}, 4.74047 \times \mu_{\mathrm{min}} \times r)
\end{equation}
and
\begin{equation}
b(r) = \mathrm{min}(v_{\mathrm{max}}, 4.74047 \times \mu_{\mathrm{max}} \times r),
\end{equation}
where the subscript $h$ denotes the properties of a {\it h-pixel}; $v_{\mathrm{min}}$ and $v_{\mathrm{max}}$ are the minimum and maximum tangential velocity limit; the factor of $4.74047$ comes from the unit conversion from arcsec\,yr$^{-1}$ to km\,s$^{-1}$ at distance $r$ in unit of pc; 
$4.74047\mu_{\mathrm{min}}r$ and $4.74047\mu_{\mathrm{max}}r$ are the tangential
velocity limits at distance $r$ arising from the proper motion limits. The
appropriate limits on the integral are found by considering both of them.

\begin{figure}
\label{fig:vtan_distribution}
\includegraphics[width=84mm]{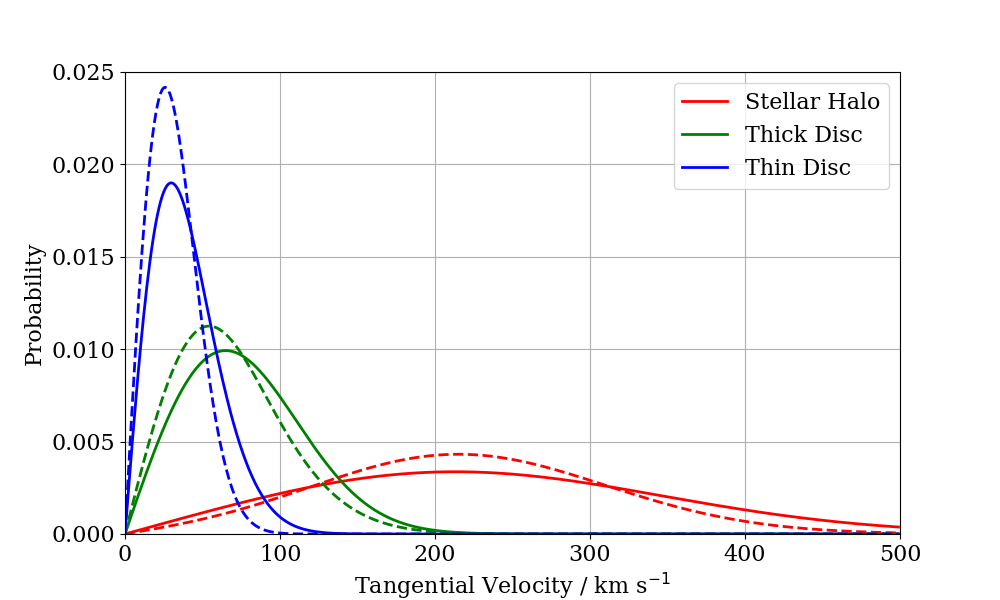}
\caption{The tangential velocity distribution of the thin disc, thick disc and 
stellar halo in the direction of the North Galactic Pole~(solid lines) and the 
Galactic Anti-Center~(dashed line) based on the kinematic information from 
Table~\ref{table:galactic_structure}.}
\end{figure}

\begin{table}
\centering
\caption{Physical properties of the Galaxy used in the Schwarzschild distribution functions. The thick disc parameters are provided for illustration purpose only.}
\begin{tabular}{ l *{3}{c} }
\hline
\hline
Parameter & Thin disc & Thick disc & Stellar Halo\\
\hline
$\langle$U$\rangle/\kmps$ & -8.62$^{a}$ & -11.0$^{c}$ & -26.0$^{c}$ \\
$\langle$V$\rangle/\kmps$ & -20.04$^{a}$ & -42.0$^{c}$ & -199.0$^{c}$ \\
$\langle$W$\rangle/\kmps$ & -7.10$^{a}$ & -12.0$^{c}$ & -12.0$^{c}$ \\
$\sigma_{U}/\kmps$ & 32.4$^{a}$ & 50.0$^{c}$ & 141.0$^{c}$ \\
$\sigma_{V}/\kmps$ & 23.0$^{a}$ & 56.0$^{c}$ & 106.0$^{c}$ \\
$\sigma_{W}/\kmps$ & 18.1$^{a}$ & 34.0$^{c}$ & 94.0$^{c}$ \\
H/pc & 250$^{b}$ & 1000$^{d}$ & $\infty$ \\
\hline
\end{tabular}
\begin{enumerate}
\itemsep0em
\item[a] \citet{2009AJ....137..266F}
\item[b] \citet{1998A&A...333..106M}
\item[c] \citet{2000AJ....119.2843C}
\item[d] \citet{2006AJ....132.1768G}
\end{enumerate}
\label{table:galactic_structure}
\end{table}

\subsection{Density Profile}
Luminous WDs near the faint limits can be several hundred parsecs from the
Galactic plane, where their space density is significantly reduced. In order to
correct for the stellar density effect on the survey volume, the density scaling
of the Galaxy has to be considered. Since the radial profile is large even when
compared to the distance of the most luminous objects, only the scaleheight,
which is perpendicular to the plane, was considered.
\subsubsection*{Thin Disc}
The thin disc employs an exponential decay law to
correct for the ``reduction in survey volume'' by the scaleheight effect. The density profile combined
with all the appropriate correction becomes
\begin{equation}
\frac{\rho(r)}{\rho_{\odot}} = \exp \left( -\frac{|r \sin(b)+z_{\odot}|}{\mathrm{H}} \right)
\end{equation}
where $|z|=r\sin{(b)}$ is the Galactic plane distance with $r$ being the 
line of sight distance, $b$ the Galactic latitude, the solar distance
from the Galactic plane is $z_{\odot} \sim 20$\,pc~\citep{2006JRASC.100..146R} and H the scaleheight.

\citet{1998A&A...333..106M} obtained a value of
$\mathrm{H}_{\mathrm{thin}} = 250$\,pc based on faint main-sequence stars. These
are likely of similar age to the WDs candidates in this work and are expected to
show a similar spatial distribution and having been subjected to the same
kinematic heating. This value is the most accepted value among works on WDLFs
although there is evidence that the scale height for faint objects is
larger\,(H06).

\subsubsection*{Stellar Halo}
The scaleheight of the halo is of order of kiloparsecs. For the depth this
work probed, the most distant objects are only a few hundred parsecs from the sun, it
is valid to assume a uniform density profile.

\subsection{Modified Volume Density Estimator}
\label{sec:Vmod}
The modified volume density estimator~(LRH15) is a variant of the maximum volume
density estimator that is generalised over a proper motion limited sample. In
order to calculate the volume available for the object, at each distance step of
the integration both the stellar density profile and discovery fractions are
considered. The total modified survey volume between r$_{\text{min}}$ and
r$_{\text{max}}$ is therefore written as
\begin{equation}
\label{eq:volume_mod}
\mathrm{V}_{\mathrm{mod}} = \Omega \int_{r_{\mathrm{min}}}^{r_{\mathrm{max}}} \frac{\rho(r)}{\rho_{\odot}} r^2 \chi(\alpha,\delta,r) \mathrm{d}r,
\end{equation}
where $\chi(\alpha,\delta,r)$ is from Equation~\ref{eq:discovery_fraction} and
the distance limits are solely determined by the photometric limits of the
survey
\begin{equation}
\label{eq:dmin_phot}
r_{\mathrm{min}} = r \times \mathrm{max} \left[ 10^{\frac{1}{5} (m_{\mathrm{min},i}-M_{i})} \right]
\end{equation}
and
\begin{equation}
\label{eq:dmax_phot}
r_{\mathrm{max}} = r \times \mathrm{min} \left[ 10^{\frac{1}{5} (m_{\mathrm{max},i}-M_{i})} \right] .
\end{equation}
The number density of a given magnitude bin is the sum of the inverse modified volume
\begin{equation}
\Phi_{k} = \sum_{i}^{N_{k}} \frac{1}{\textrm{V}_{\textrm{mod, i}}},
\end{equation}
for N$_{k}$ objects in the $k$-th bin. The uncertainty of each star's
contribution is assumed to follow Poisson statistics. The sum of all errors in
quadrature within a luminosity bin is therefore written as
\begin{equation}
\sigma_{k} = \left[ \sum_{i=1}^{N_{k}} \left( \frac{1}{\textrm{V}_{\textrm{mod}}} \right)^{2} \right]^{1/2} .
\end{equation}

\subsection{Voronoi Tessellation}

A Voronoi tessellation is made by partitioning a plane with $n$ points into $n$ convex polygons such that each polygon contains one point. Any position in a given polygon~(cell) is closer to its generating point than to any other points. For use in astronomy, such a tessellation has to be done on a spherical surface~(two-sphere).

In this work, the tessellation is constructed with the {\sc SciPy} package spatial.SphericalVoronoi, where each polygon is given a unique ID that is combined with the vertices to form a dictionary. The areas are calculated by first decomposing the polygons into spherical triangles with the generating points and their vertices~\citep{tyler.reddy.2015.13688} and then by using L'Huilier's Theorem to find the spherical excess. For a unit-sphere, the spherical excess is equal to the solid angle of the triangle. The sum of the constituent spherical triangles provides the solid angle of each cell. See Section 2 of L17 for detailed description.

\subsection{Cell Properties}
\label{sec:cell_properties}
For a Voronoi cell $j$, the properties of the cell are assumed to be represented by generating source $i$. Both $i$ and $j$ are indexed from $1$ to $\mathcal{N}$, but since each source has to be tested for observability in each cell to calculate the maximum volume, $i$ and $j$ cannot be contracted to a single index. Furthermore, the cells do not need to be defined by only the sources of interest. Arbitrary points can be used for tessellation such that $i$ and $j$ will not have a one-to-one mapping. The epoch of the measurement is labelled by $k$. In this work, we use the full catalogue with $14,598$ sources to generate the Voronoi tessellation, and analysis were performed using this fixed set of cells (See Table~\ref{table:catalogue} for the catalogue of these sources, and Table~\ref{table:epoch} for the epoch information).

\subsection{Voronoi V$_{\mathrm{max}}$}
\label{sec:volume_voronoi}
In order to incorporate the Voronoi tessellation into the modified volume method, two minor adjustments are required to apply to the volume integral -- (1) the lower proper motion limit; and (2) the area element $\Omega_j$ in Equation~\ref{eq:vmax}.

\subsubsection*{Lower Proper Motion Limit}
\begin{multline}
\label{eq:vmax}
\mathrm{V}_{\mathrm{max}} = \sum_j \Omega_j \int_{r_{\mathrm{min,j}}}^{r_{\mathrm{max,j}}} \dfrac{\rho(r)}{\rho_{\odot}} \times r^2 \\
\hfill \times \left[ \int_{a(r)}^{b(r)} P_{h(j)}(v_{\mathrm{T}}) \, \mathrm{d}v_{\mathrm{T}} \right] \mathrm{d}r
\end{multline}
where $\dfrac{\rho(r)}{\rho_{\odot}}$ is the density normalized by that at the solar neighbourhood, $P_{h(j)}$ is the tangential velocity distribution, $h(j)$ denotes the h-pixel mapped from cell $j$ with area $\Omega_j$, $v_{\mathrm{T}}$ is the tangential velocity, $r_{\mathrm{min}}$ and $r_{\mathrm{max}}$ are the minimum and maximum photometric distances, and $\sigma_{\mu}(r)$ is the proper motion uncertainty as a function of the distance to the source. Consequentially, at each step of the integration, the $\sigma_{\mu}$ has to be recomputed~(L17, necessary epoch informations can be found in \ref{table:epoch}). With a set of catalogued observational data, new interstellar reddening has to be applied at the new distance before the ``new observed flux'' is converted into instrumental flux by using the given zero-points. A set of new photometric and astrometric uncertainties can then be recomputed based on the instrumental flux, epoch sky brightness, dark current and read noise. The uncertainties are checked against the desired limits in order to identify the distance limit for the volume integration. The lower tangential velocity limit in the inner integral, $a(r)$, is
\begin{equation}
\label{eq:inner_limit}
a(r) = \mathrm{max}\left[v_{\mathrm{min}}, 4.74047\times s \times\sigma_{\mu}(r) \times r \right]
\end{equation}
where $v_{\mathrm{min}}$ is the global lower tangential velocity limit and $s$ is the significance of the proper motions, which is $7.5$ in this work.

\subsubsection{Voronoi Cell Area}
\begin{figure}
\includegraphics[width=84mm]{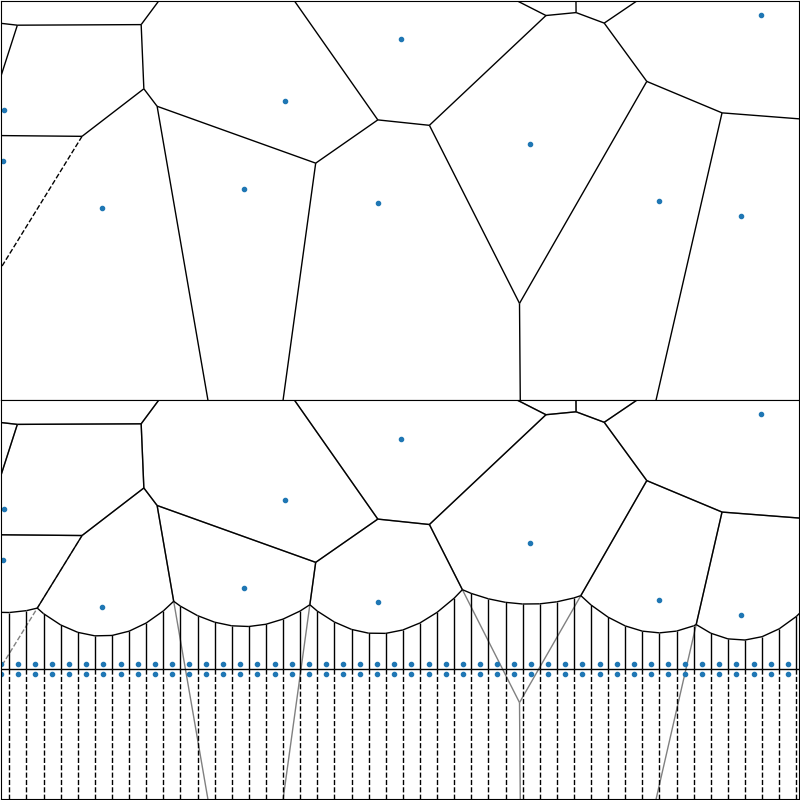}
\caption{Top: A simple illustration of Voronoi cells at the survey boundaries. Bottom: Voronoi cells after artificial points added. Grey lines shows the original cell boundaries. The areas generated from the artificial points are added to the host cell, which is approximated by assigning the artificial points to the host cell. In the observed sample, the number ratio between the cells containing a genuine object and the artificial cells is much larger than it is shown here for illustration purpose.}
\label{fig:cell_boundaries}
\end{figure}

In the framework of L17, the simulated data is an all sky survey with no spatial selection criteria. However, in any given survey or analysis, there are usually spatial limits (e.g.,~selection or limits on right ascension and declination; or leaving out dense regions to avoid confusion). When such selections are necessary, since the Voronoi cells constructed with the data always cover the entire sphere (i.e.,~the total solid angle is $4\pi$), it is necessary to add artificial points to the set of data in order to define the set of Voronoi cells that carry the appropriate areas. In order to align the artificial cells boundaries with the selection borders, tightly spaced points along two rings at equidistance from the border are required. In this work, we add $21,600$ points on each ring which would be equivalent to a spacing of $1\arcmin$ at the equator. The spacing between the ring and the border is $3\arcsec$. By identifying where the artificial points belong in the original Voronoi cells (hereafter the bounding cells), the areas of the Voronoi cells generated from the respective artificial points are added to the new areas of the bounding cells (see Fig.~\ref{fig:cell_boundaries}). It is trivial to add points along a single coordinate axis. However, when the area within the $20\degr$ radius from the galactic centre is removed from our survey, the positions that trace two rings on either side at eqidistance from the selection boundary are not trivial to calculate. It is much simpler to define a dummy coordinate system such that the Galactic Centre is located at the Pole. In such configuration, the rings can be defined by a single coordinate axis. This can be done by rotating the Galactic Coordinates by $90\degr$ along the vector joining the centre of the celestial sphere to $(l,b) = (90,0)$. We use the Euler-Rodrigues formula for this purpose. 

There is one caveat, the lines joining the Voronoi vertices are great circle lines. However, many selections, for example lines of equal-declination, are traced by small circles, so the area of any Voronoi cells constructed this way are only approximations, but they are only offsets by negligible amounts.

\subsection{Interstellar reddening}
Interstellar reddening has small effect in determining the distance and bolometric magnitude of an individual object. However, it causes a change in the shape of a WDLF when a large sample is considered. When WDs cool down, they turn red until they reach $\sim$$6,000$\,K beyond which they start to turn blue due to H2CIA. Therefore, the hot and cool WDs require larger corrections than the warm ones. Without extinction correction, the bright end of the WDLF will have a larger gradient~(more positive), while the faint end will have a smaller~(more negative) gradient. In order
to correct for the interstellar reddening, Equations~\ref{eq:dmin_phot}
and~\ref{eq:dmax_phot} have to be modified to
\begin{equation}
\label{eq:dmin_red_phot}
r_{\mathrm{min}} = r \times \mathrm{max} \left[ 10^{\frac{1}{5} (m_{\mathrm{min},i}-M_{i}-A_{i}(r)+A_{i}(r_{\mathrm{min}}))} \right]
\end{equation}
and
\begin{equation}
\label{eq:dmax_red_phot}
r_{\mathrm{max}} = r \times \mathrm{min} \left[ 10^{\frac{1}{5} (m_{\mathrm{max},i}-M_{i}-A_{i}(r)+A_{i}(r_{\mathrm{max}}))} \right] .
\end{equation}

\section{White Dwarf Luminosity Functions}
\label{sec:wdlf_sec}
In addition to applying all the selection criteria discussed in Section 2, finder charts were inspected to remove spurious objects. The survey contains $14,598$ WD candidates with $v_{\textrm{tan}} > 40\kmps$ and proper motion at $7.5\sigma$ significance. However, a proportion of them do not enter any of the analysis due to the stringent velocity and photometric parallax quality selection in deriving WDLFs that are representative of the disc~(low velocity sample) and halo~(high velocity sample).

\subsection{WDLFs combining two Atmosphere Models}
To limit contaminations, the maximum goodness-of-fit reduced chi-squared of the photometric parallax~($\chi^{2}_{\mu}$) is set at $10$ above $6000$\,K and at $2$ below that. The smaller tolerance comes from the ``blue hook'' of the WD cooling sequence where spurious objects~(e.g.,~high proper motion subdwarfs) are much more likely to be fitted as WDs. The mixed hydrogen-helium atmosphere model is only available below $12,000$\,K~($\sim$$11.5\magnitude$). Above this there is little difference in the models and only DA is considered in constructing the luminosity function. Because of the lack of available DB models above $12,000$\,K, objects in the range of $10,000-12,000$\,K tend to have poor goodness-of-fit. To avoid this systematic bias, objects are divided into $3$ groups where the latter 2 are summed with appropriate weightings to give the total WDLFs :
\begin{itemize}
\item[i.] Objects with best fit DA temperature above $10,000$\,K;
\item[ii.] Objects with best fit DA temperature below $10,000$\,K and have good DA fit;
\item[iii.]  Objects with best fit DA temperature below $10,000$\,K and have good DB fit\footnote{A WD can be at two different temperature/magnitude bins with DA and DB models}.
\end{itemize}
Objects in (i) is unit-weighted; for those in (ii) and (iii), they are weighted by the using the reduced chi-squared value, $\chi^{2}_{\mu}$, of the photometric parallax. The probability of an object being a DA and DB are $P_{\mathcal{A}} \propto \exp(-0.5 \chi^{2}_{\mu,\mathcal{A}})$ and $P_{\mathcal{B}} \propto \exp(-0.5 \chi^{2}_{\mu,\mathcal{B}})$ respectively. The weights of objects being DA and DB are the ratio of the two probabilities. The total luminosity function is the weighted sum of the inverse maximum volume.

\subsection{WDLF of the Low Velocity Sample in the Solar Neighbourhood}
\label{sec:wdlf_solar_neighbourhood}

\begin{figure}
\includegraphics[width=84mm]{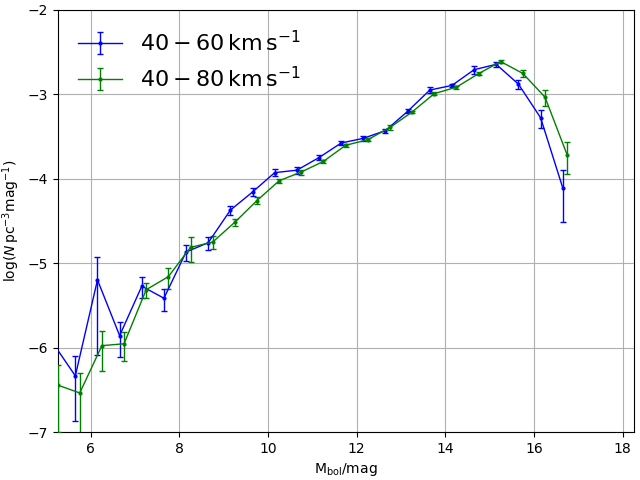}
\caption{WDLF of the low velocity samples in the solar neighbourhood. The $40-80\kmps$ line is shifted by $0.1\magnitude$ for easier visual comparison. The two samples agree well with each other.}
\label{fig:wdlf_solar_neighbourhood}
\end{figure}

The WDLFs of the low velocity sample~(hereafter, disc), are shown in Fig.~\ref{fig:wdlf_solar_neighbourhood}. In the $40-60$ and $40-80\kmps$ samples, there are $6,495$ and $9,561$ WD candidates and the integrated WD densities are $5.314 \pm 0.487 \times 10^{-3}$ and $5.657 \pm 0.416 \times 10^{-3}$\,pc$^{-3}$ respectively, where the corresponding $\left\langle \mathrm{V}/\mathrm{V}_{\mathrm{max}}\right\rangle$s are $0.547\pm0.004$ and $0.556\pm0.003$. Since the cooling time for DA with $\log(g) = 8.0$ to reach $16.0$, $16.5$, $17.0$, $17.5$ and $18.0$\,mag are $9.36$, $10.39$, $11.16$, $11.84$ and $12.49\gyr$ respectively, and $8.61$, $9.46$, $10.27$, $11.07$ and $11.88\gyr$ for DB; the faintest objects are most likely coming from the low velocity tail of the thick disc kinematic distribution. An alternative explanation is that they are low mass WD that have a higher cooling rate and lower surface gravity: at $\log(g) = 7.0$, the cooling ages for DA drop to $3.44$, $4.24$, $6.33$, $9.02$ and $11.62\gyr$. However, this is inconsistent with the assumption of a fixed surface gravity $\log(g) = 8.0$ in our analysis. In the $25$\,pc volume limited sample from \citet{2016MNRAS.462.2295H}, at the $17.5\magnitude$ bin, there are a massive DB, with $\log(g) = 9.0$,  belonging to a widely separated double degenerate system and a DA presumed to have $\log(g) = 8.0$. SED fitting with 5-band broadband photometry cannot reliably fit the surface gravity or surface hydrogen/helium ratio as free parameters. It will only be possible for a large sky area survey to expand the fitting parameter space with, for example, the future Gaia data release where parallax and low resolution spectra would be available for most of the nearby sources.

The smaller the range of tangential velocities, the fewer contaminants from the disc main sequence stars. However, the WDLF would become more model dependent on the Galactic model and more sensitive to the completeness corrections.

\subsection{WDLF of the High Velocity Sample in the Solar Neighbourhood}

\begin{figure}
\includegraphics[width=84mm]{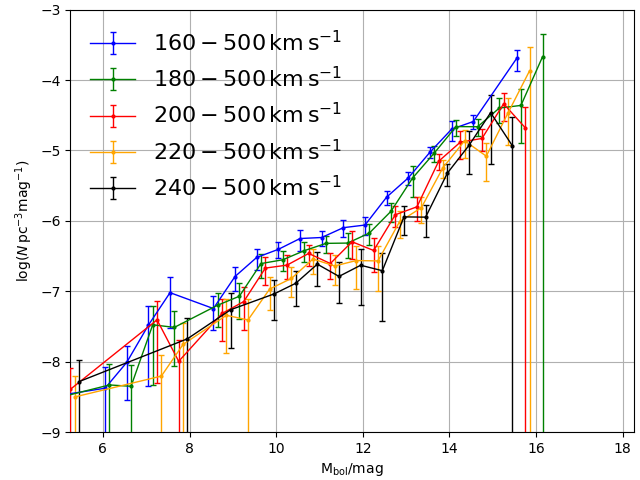}
\caption{WDLF of the high velocity sample in the solar neighbourhood. The WDLF with a selection of $200-500\kmps$ is located at the correct magnitude. Each successive decrement and increment in the lower tangential velocity limit shifts the WDLFs by $-0.1\magnitude$ and $+0.1\magnitude$ respectively for easier visual comparison. All WDLFs agree well with each other up to $15.25\magnitude$ when the larger lower-tangential velocity limit removes the faintest objects.}
\label{fig:wdlf_halo}
\end{figure}

The high velocity samples contain $1.334 \pm 0.420 \times 10^{-4}$, $1.798 \pm 1.487 \times 10^{-4}$, $5.291 \pm 2.717 \times 10^{-5}$, $1.006 \pm 0.950 \times 10^{-4}$ and $3.296 \pm 2.849 \times 10^{-5}$\,pc$^{-3}$ for the tangential velocity selection between $160$, $180$, $200$, $220$ \& $240\kmps$ and $500\kmps$. The five samples have $\left\langle \mathrm{V}/\mathrm{V}_{\mathrm{max}}\right\rangle$ at $0.427\pm0.016$, $0.447\pm0.020$, $0.459\pm0.024$, $0.434\pm0.031$ and $0.432\pm0.039$. The decrease in the number density comes from the lack of the faintest candidates in the higher velocity samples; the five WDLFs agree with each other at the brighter end is a good indication that the samples are properly normalized~(Fig.~\ref{fig:wdlf_halo}). The $160$ \& $180\kmps$ samples are contaminated by a non-negligible amount of thick disc WDs, the faintest bins in the $160\kmps$ sample are most likely coming from the disc. The $200\kmps$ sample should be the lowest reliable tangential velocity cut for testing the sample as from a halo population~(RH11, M17). However, one has to be cautious when selecting sub sample as halo candidates as it is very likely we are looking at the tail of the thick disc distribution~\citep{2001Sci...292..698O, 2001ApJ...559..942R}. It appears that the down turn of the halo WDLF is still out of reach of Pan--STARRS 1.

\subsection{Data available online}
Machine readable text files are available online as supplementary materials to generate Figure~\ref{fig:wdlf_solar_neighbourhood} and \ref{fig:wdlf_halo}. See Table~\ref{table:wdlf} for the description.

\section{Comparison with Previous Works}

There are three works on WDLFs in the past $20$ years employing large sky area photometric surveys: the H06, RH11 and M17. The former two rely on photographic plates so they are much more limited by the astrometry in the faint end. In M17, the combination of SDSS, the Bok $90$-inch telescope at the Steward Observatory and the $1.3$\,m telescope at Flagstaff Station, USNO has enabled unprecedented photometric and astrometric quality for work on WDLF to date. Its strengths are both the low proper motion uncertainty and survey depth; in comparison, 3$\upi$ Survey strength is the rapid re-imaging and the full visible sky at Haleakal$\bar{\mathrm{a}}$ which gives a footprint area $11.4$ times\footnote{This ratio is between the areas used in the respective works, rather than the ratio of the entire surveys.} larger than that in M17. We believe the M17 WDLFs should be considered as the best reference available at this moment. However, when drawing comparisons, it is unclear whether it has probed sufficiently far away that the study has reached some local Galactic structures of over- or under-density.

\subsection{Low Velocity Sample/disc(s)}

\begin{figure}
\includegraphics[width=84mm]{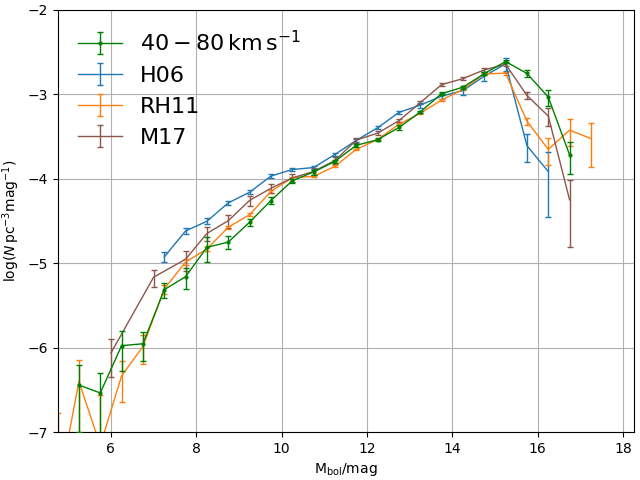}
\caption{Comparison of WDLFs of the low velocity sample in H06, RH11, M17 and this work. This work is reporting a similar integrated number density to all previous works.}
\label{fig:wdlf_solar_neighbourhood_compared}
\end{figure}

As shown in Fig.~\ref{fig:wdlf_solar_neighbourhood_compared}, our WDLF has the same general shape compared to the past works, which is expected as it takes a significantly different Galactic density profile or star formation history in order for the shape to vary noticeably. This work has very similar density to all previous works, despite the use of a new generalized maximum volume method. We note that H06 and M17 have similar footprints; and the footprint in this work is similar to that in RH11. The density differences in local Galactic structures or an evolving scaleheight, instead of a fixed $250$\,pc, may attribute to some of the discrepancies. From H06, it is understood that a fixed scaleheight is not the most appropriate assumption in studying the disc sample: fainter populations follow larger scaleheights, due to kinematic heating of the discs. The smaller footprint area and less coverage near the Galactic plane in H06 and M17 means that the variations in density is likely to be smaller. The large footprint area at greater depth in this work as compared to RH11 could have amplified the effect. In the $4$ works shown in Fig.~\ref{fig:wdlf_solar_neighbourhood_compared}, we suspect that H06, with the smallest sample volume, is displaying the feature at the bright end of the WDLF that was understood as an enhanced star formation from the $25$\,pc sample~\citet{2016MNRAS.462.2295H}. While in the M17, which is essentially a deeper version of H06, such enhanced density is not shown; in RH11 and this work, the footprint areas are a few times larger, small scale~(hundreds of squared degrees) features are likely to be averaged out. The different atmosphere models adopted by the four works can also contribute to the discrepancies. The bumps at $\sim$$10$ and $\sim$$12\magnitude$ appear in all $4$ works is evidence that they are genuine features of recent star burst~($\sim$$1\gyr$). This feature is more prominent for a more stringent volume-limited sample~\citep{2017ASPC..509...59O}. The precise time of the star burst can only be revealed by a proper SFH analysis. The integrated number density $5.657 \pm 0.416 \times 10^{-3}$\,pc$^{-3}$ of this work is in very good agreement with the $5.5 \pm 0.1 \times 10^{-3}$\,pc$^{-3}$ from M17; and the $4.6 \pm 0.5 \times 10^{-3}$\,pc$^{-3}$ by H06.

\subsection{High Velocity Sample/Halo}

\begin{figure}
\includegraphics[width=84mm]{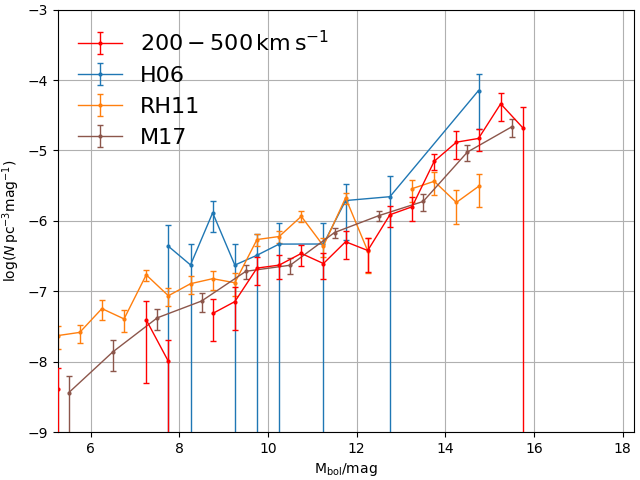}
\caption{Comparisons of WDLFs of the high velocity sample in H06, RH11, M17 and this work. This work has a lower density in the range of $8-13\magnitude$, but it still agrees to within $1\sigma$ combined uncertainties. The integrated number densities are all within $1\sigma$ confidence from each other, except for RH11.}
\label{fig:wdlf_halo_compared}
\end{figure}

The WDLF ($200-500\kmps$) of the high velocity sample agrees well with previous works~(Fig.~\ref{fig:wdlf_halo_compared}). The integrated density at $5.291 \pm 2.717 \times 10^{-5}$\,pc$^{-3}$ is slightly higher than $4.0 \times 10^{-5}$\,pc$^{-3}$ and $3.5 \pm 0.7 \times 10^{-5}$\,pc$^{-3}$ by H06 and M17 respectively, but they are within $1\sigma$ confidence limit from each other; and it is well under $1.9 \times 10^{-4}$\,pc$^{-3}$ reported by the effective volume method~(RH11). The disc-to-halo ratio in this work is $107$, which is about $30\%$ smaller than the $157$ found in M17; very similar to H06's value at $115$. The most appropriate comparison from RH11 is the ratio between the sum of the densities of the discs and that of the halo found from the effective volume methods, at $19.7$. However, it is worth noting the different faint limits each WDLF probes, the lack of data in the highest density bin, which is most likely in the range $16-18\magnitude$, bias the disc-to-halo ratio significantly.

\subsection{Data available online}
Machine readable text files are available online as supplementary materials to generate Figure~\ref{fig:wdlf_solar_neighbourhood_compared} and \ref{fig:wdlf_halo_compared}. See Table~\ref{table:wdlf_compared} for the description.

\section{Summary and Future Work}

We have applied the newest V$_{\mathrm{max}}$ method from L17, which formally propagates instrumental noise from individual epoch to proper motions uncertainties, to derive the disc and halo WDLFs from the Pan--STARRS~1 3$\pi$ Survey. The number densities are found to be $5.657 \pm 0.416 \times 10^{-3}$\,pc$^{-3}$ and $5.291 \pm 2.717 \times 10^{-5}$\,pc$^{-3}$ respectively. Both results are consistent with previous results from studies of a similar kind.

In order to study the Galactic components independently, a rigorous statistical method has to be devised~(e.g.,~extending on RH11, \citealt{2017ASPC..509...25L}) in order to derive the star formation history of the individual components through the inversion of the respective WDLFs. The WD candidates from the Gaia DR2 selected by parallax~(e.g.,~\citealt{2018MNRAS.480.3942H, 2018arXiv180703315G}) as opposed to by proper motion and the subsequent releases will provide an order of magnitude more WDs with full 5 astrometric solutions and low resolution spectra~(DR3+) will potentially shed new light to the understanding of this, currently, elusive population; better understanding to the kinematics and density profiles can derive more accurate WDLFs. In the subsequent work, we will apply a similar selection and analysis on the Gaia data, which will show the definite solution in the next couple of decades. It is the dawn of WD science in this coming era with multiple large sky area photometric and spectrometric surveys coming online.

\section*{Acknowledgments}
The Pan--STARRS 1 Surveys\,(PS1) have been made possible through contributions
of the Institute for Astronomy, the University of Hawaii, the Pan--STARRS
Project Office, the Max-Planck Society and its participating institutes, the Max
Planck Institute for Astronomy, Heidelberg and the Max Planck Institute for
Extraterrestrial Physics, Garching, The Johns Hopkins University, Durham
University, the University of Edinburgh, Queen's University Belfast, the
Harvard-Smithsonian Center for Astrophysics, the Las Cumbres Observatory Global
Telescope Network Incorporated, the National Central University of Taiwan, the
Space Telescope Science Institute, the National Aeronautics and Space
Administration under Grant No. NNX08AR22G issued through the Planetary Science
Division of the NASA Science Mission Directorate, the National Science
Foundation under Grant No. AST-1238877, the University of Maryland, and Eotvos
Lorand University\,(ELTE).

The star galaxy separator of this work is based on observations obtained with
MegaPrime/MegaCam, a joint project of CFHT and CEA/DAPNIA, at the
Canada-France-Hawaii Telescope~(CFHT) which is operated by the National Research
Council~(NRC) of Canada, the Institut National des Sciences de l'Univers of the
Centre National de la Recherche Scientifique~(CNRS) of France, and the
University of Hawaii. This research used the facilities of the Canadian
Astronomy Data Centre operated by the National Research Council of Canada with
the support of the Canadian Space Agency. CFHTLenS data processing was made
possible thanks to significant computing support from the NSERC Research Tools
and Instruments grant program.

We thank the PS1 Builders and PS1 operations staff for construction and
operation of the PS1 system and access to the data products provided.

ML acknowledges financial support from a UK STFC. ML also acknowledges computing resources available on the
Stacpolly and Cuillin clusters at the IfA~(Edinburgh) and on the cluster
at the ARI~(Liverpool JM).

\bibliography{2018_ps_wdlf}

\appendix
\section{Supplementary Materials}
The following tables describe the content of the supplementary material available online. In Table~\ref{table:catalogue} and \ref{table:epoch}, the joint Catalogue ID and Object ID form an unique ID to map the epoch measurements to the source. However, this is not an unique ID over different processing versions and data releases. There are not direct mappings between the sources in this catalogue~(PV2) and the public releases DR1 and DR2. Table~\ref{table:epoch} is divided into two compressed files, first one contains all measurements with R.A. between $0\degr$ and $180\degr$; and the second one with $180\degr$ to $360\degr$. 
\begin{table*}
\centering
\caption{Description of the full catalogue used to generate the Voronoi Tessellation containing $14,598$ sources. A number of these sources were only used to model the survey properties but were not directly used in computing the WDLFs. See Section~\ref{sec:wdlf_sec} for the selection criteria for the various WDLFs. The joint Catalogue ID and Object ID forms an unique ID, however, this is not unique over different processing versions and data releases.}
\label{table:catalogue}
\begin{tabular}{ll}
Column & Description \\ \hline \hline 
1 & Right Ascension (epoch at the Mean Epoch and equinox at 2000.0)\\
2 & Declination (epoch at the Mean Epoch and equinox at 2000.0)\\
3 & Catalogue ID\\
4 & Object ID\\
5 & Mean Epoch (number of seconds since 1 January 1970)\\
6 & Original Proper Motion in the direction of R.A. $(''\,\mathrm{yr}^{-1})$\\
7 & Original Proper Motion Uncertainty in the direction of R.A. $(''\,\mathrm{yr}^{-1})$\\
8 & Original Proper Motion in the direction of Dec. $(''\,\mathrm{yr}^{-1})$\\
9 & Original Proper Motion Uncertainty in the direction of Dec. $(''\,\mathrm{yr}^{-1})$\\
10 & Original Chi-squared Value in Proper Motion Solution\\
11 & Recomputed Proper Motion in the direction of R.A. $(''\,\mathrm{yr}^{-1}$)\\
12 & Recomputed Proper Motion in the direction of Dec. $(''\,\mathrm{yr}^{-1}$)\\
13 & Recomputed in Proper Motion Uncertainty (same in the two directions)\\
14 & Recomputed Chi-squared Value in the direction of R.A.\\
15 & Recomputed Chi-squared Value in the direction of Dec.\\
16 & g$_{\mathrm{p1}}$ PV2 magnitude (mag)\\
17 & $\sigma_{\mathrm{g}_{\mathrm{p1}}}$ PV2 magnitude (mag)\\
18 & r$_{\mathrm{p1}}$ PV2 magnitude (mag)\\
19 & $\sigma_{\mathrm{r}_{\mathrm{p1}}}$ PV2 magnitude (mag)\\
20 & i$_{\mathrm{p1}}$ PV2 magnitude (mag)\\
21 & $\sigma_{\mathrm{i}_{\mathrm{p1}}}$ PV2 magnitude (mag)\\
22 & z$_{\mathrm{p1}}$ PV2 magnitude (mag)\\
23 & $\sigma_{\mathrm{z}_{\mathrm{p1}}}$ PV2 magnitude (mag)\\
24 & y$_{\mathrm{p1}}$ PV2 magnitude (mag)\\
25 & $\sigma_{\mathrm{y}_{\mathrm{p1}}}$ PV2 magnitude (mag)\\
26 & DA Photometric Distance (pc)\\
27 & DA Photometric Distance at 1 sigma lower limit (pc)\\
28 & DA Photometric Distance at 1 sigma upper limit (pc)\\
29 & DA Photometric Temperature (K)\\
30 & DA Photometric Temperature at 1 sigma lower limit (K)\\
31 & DA Photometric Temperature at 1 sigma upper limit (K)\\
32 & DA Photometric Absolute Bolometric Magnitude (mag)\\
33 & DA Photometric Absolute Bolometric Magnitude at 1 sigma lower limit (mag)\\
34 & DA Photometric Absolute Bolometric Magnitude at 1 sigma upper limit (mag)\\
35 & DA Photometric Solutions Chi-squared Value\\
36 & DB Photometric Distance (pc)\\
37 & DB Photometric Distance at 1 sigma lower limit (pc)\\
38 & DB Photometric Distance at 1 sigma upper limit (pc)\\
39 & DB Photometric Temperature (K)\\
40 & DB Photometric Temperature at 1 sigma lower limit (K)\\
41 & DB Photometric Temperature at 1 sigma upper limit (K)\\
42 & DB Photometric Absolute Bolometric Magnitude (mag)\\
43 & DB Photometric Absolute Bolometric Magnitude at 1 sigma lower limit (mag)\\
44 & DB Photometric Absolute Bolometric Magnitude at 1 sigma upper limit (mag)\\
45 & DB Photometric Solutions Chi-squared Value \\ \hline
\end{tabular}
\end{table*}

\begin{table}
\centering
\caption{Description of the table containing all the necessary epoch information to model the survey properties during the maximum volume integration over the Voronoi tessellation cells. The conversion between the instrumental magnitude and the relative magnitude can be calculated by $\mathrm{mag}_{\mathrm{rel}} = \mathrm{mag}_{\mathrm{inst}} - 25.0 + \texttt{C\_LAM} \times 0.001 + \texttt{K} \times (\texttt{AIRMASS} - 1.0) - \texttt{M\_CAL}$ where \texttt{PHOTOCODE} $10001-10076$ correspond to g band photometry, $10101-10176$ correspond to r band, $10201-10276$ correspond to i band, $10301-10376$ correspond to z band and $10401-10476$ correspond to y band; the values of \texttt{C\_LAM} in the grizy bands are 24563, 24750, 24611, 24250 and 23320 respectively; and the values of \texttt{K} in the grizy bands are -0.147, -0.085, -0.044, -0.033 and -0.073 respectively.}
\label{table:epoch}
\begin{tabular}{ll}
Column & Description \\ \hline \hline 
1 & Right Ascension\\
2 & Declination\\
3 & Instrumental Magnitude (mag)\\
4 & Instrumental Magnitude Uncertainty (mag)\\
5 & \texttt{M\_CAL} (mag)\\
6 & Exposure time (s)\\
7 & Airmass\\
8 & Sky background flux (weighted PSF flux)\\
9 & Epoch (number of seconds since 1 January 1970)\\
10 & Object ID\\
11 & Catalogue ID\\
12 & \texttt{PHOTCODE} - filter and detector chip ID\\ \hline
\end{tabular}
\end{table}

\begin{table}
\centering
\caption{Description of the machine readable text files to generate Figure~\ref{fig:wdlf_solar_neighbourhood} and \ref{fig:wdlf_halo}. Bright magnitude solutions are not reliable without UV photometry and are not shown in the figures, they are only included in the text files as part of the complete set of solutions. The number of sources are not always integers because they come from the weighted sum of DA and DB WDLFs.}
\label{table:wdlf}
\begin{tabular}{ll}
Column & Description \\ \hline \hline
1 & Bolometric Magnitude (mag)\\
2 & Number Density $n$ (N pc$^{-3}$)\\
3 & $\sigma_{n}$ (N pc$^{-3}$)\\
4 & Number of sources \\
\hline
\end{tabular}
\end{table}

\begin{table}
\centering
\caption{Description of the machine readable text files to generate Figure~\ref{fig:wdlf_solar_neighbourhood_compared} and \ref{fig:wdlf_halo_compared}. Bright magnitude solutions are not reliable without UV photometry and are not shown in the figures, they are only included in the text files as part of the complete set of solutions. The number of sources are not always integers because they come from the weighted sum of DA and DB WDLFs.}
\label{table:wdlf_compared}
\begin{tabular}{ll}
Column & Description \\ \hline \hline
1 & Bolometric Magnitude (mag)\\
2 & Number Density $n$ (N pc$^{-3}$)\\
3 & $\sigma_{n}$ (N pc$^{-3}$)\\
\hline
\end{tabular}
\end{table}

\label{lastpage}

\end{document}